\documentclass{aa}  

\usepackage{graphicx}
\usepackage{graphics}
\usepackage{natbib}
\usepackage{adjustbox}
\usepackage{amsmath}
\usepackage{xcolor}
\usepackage{float}
\usepackage{caption}
\bibpunct{(}{)}{;}{a}{}{,} 

\usepackage{txfonts}
\usepackage{amsmath}
\begin{document}

\newcommand{\sigmahtwo}{$\Sigma_{\rm H_2}$}
\newcommand{\sigmahi}{$\Sigma_{\rm HI}$}
\newcommand{\sigmasfr}{$\Sigma_{\rm SFR}$}
\newcommand{\hi}{H{\sc i}}
\newcommand{\micron}{$\mu$m}
\newcommand{\taudepl}{$\tau_{\rm depl}$}
\newcommand{\msunpc}{${\rm M}_\odot\,{\rm pc}^{-2}$}

\title{
The ISM scaling relations in DustPedia\thanks{DustPedia is a project funded by the EU under the heading 
`Exploitation of space science and exploration data'. 
It has the primary goal of exploiting existing data in the 
\textit{Herschel} Space Observatory and Planck Telescope databases. 
} late-type galaxies: \\
A benchmark study for the Local Universe
}
\author{V. Casasola$^{1,2}$, 
S. Bianchi$^{2}$, 
P.~De~Vis$^{3}$,
L.~Magrini$^{2}$, 
E.~Corbelli$^{2}$, 
C.~J.~R.~Clark$^{4}$,
J.~Fritz$^{5}$,
A.~Nersesian$^{6,7,8}$,
S.~Viaene$^{8}$,
M.~Baes$^{8}$, 
L.~P. Cassar\`a$^{9,6}$, 
J.~Davies$^{3}$, 
I.~De~Looze$^{10,8}$, 
W.~Dobbels$^{8}$,
M.~Galametz$^{11}$,
F.~Galliano$^{11}$, 
A.~P.~Jones$^{12}$,     
S.~C.~Madden$^{11}$,
A.~V.~Mosenkov$^{13,14}$, 
A.~Tr\v cka$^{8}$, 
E. Xilouris$^{6}$ 
}
\institute{
$^{1}$ INAF -- Istituto di Radioastronomia, Via P. Gobetti 101, 40129, Bologna, Italy\\
\email{viviana.casasola@inaf.it}  \\
$^{2}$ INAF -- Osservatorio Astrofisico di Arcetri, Largo E. Fermi 5, 50125, Firenze, Italy\\
$^{3}$ School of Physics and Astronomy, Cardiff University, The Parade, Cardiff CF24 3AA, UK \\
$^{4}$ Space Telescope Science Institute, 3700 San Martin Drive, Baltimore, Maryland, 21218, USA \\
$^{5}$ Instituto de Radioastronom\'\i a y Astrof\'\i sica, UNAM, Campus Morelia, A.P. 3-72, C.P. 58089, Mexico\\
$^{6}$ National Observatory of Athens, Institute for Astronomy, Astrophysics, Space Applications and Remote Sensing, Ioannou Metaxa and Vasileos Pavlou GR-15236, Athens, Greece \\
$^{7}$ Department of Astrophysics, Astronomy \& Mechanics, Faculty of Physics, University of Athens, Panepistimiopolis, GR-15784 Zografos, Athens, Greece \\
$^{8}$ Sterrenkundig Observatorium Universiteit Gent, Krijgslaan 281 S9, B-9000 Gent, Belgium \\ 
$^{9}$ INAF -- Istituto di Astrofisica Spaziale e Fisica cosmica,  Via A. Corti 12, 20133, Milano, Italy \\ 
$^{10}$ Department of Physics and Astronomy, University College London, Gower Street, London WC1E 6BT, UK\\
$^{11}$ Laboratoire AIM, CEA/DSM - CNRS - Universit\'e Paris Diderot, IRFU/Service d'Astrophysique, CEA Saclay, 91191, Gif-sur- Yvette, France \\
$^{12}$ Institut d'Astrophysique Spatiale, CNRS, Univ. Paris-Sud, Universit\'e Paris-Saclay,  B\^{a}t. 121, 91405, Orsay Cedex, France \\
$^{13}$ Central Astronomical Observatory of RAS, Pulkovskoye Chaussee 65/1, 196140, St. Petersburg, Russia\\
$^{14}$ St. Petersburg State University, Universitetskij Pr. 28, 198504, St. Petersburg, Stary Peterhof, Russia\\
}
\date{Received ; accepted}

\titlerunning{ISM in DustPedia late-type galaxies}
\authorrunning{V.~Casasola et al.}

\abstract
{}
{The purpose of this work is the characterization of the main scaling relations between all of the interstellar medium (ISM) components, namely dust,  
atomic, molecular, and total gas, and gas-phase metallicity, as well as
other
galaxy properties, such as stellar mass (M$_{\rm star}$) and galaxy morphology, for late-type galaxies in the Local Universe.  }
{
This study was performed by extracting late-type galaxies from the entire DustPedia sample and by exploiting the large and homogeneous dataset available thanks to the DustPedia project.   
The sample consists of 436 galaxies with morphological stage spanning from T~=~1 to 10, 
M$_{\rm star}$ from  $6 \times 10^7$ to $3 \times 10^{11}$~M$_\odot$, star formation rate from $6 \times 10^{-4}$ to 60~M$_\odot$~yr$^{-1}$, 
and oxygen abundance 
from 12~+~log(O/H)~=~8 to 9.5. 
Molecular and atomic gas data were collected from the literature and properly homogenized.
All the masses involved in our analysis refer to the values within the optical disks of galaxies. 
The scaling relations involving the molecular gas are studied by assuming both a constant and a metallicity-dependent CO-to-H$_2$ conversion factor ($X_{\rm CO}$).
The analysis was performed by means of the survival analysis technique, 
in order to properly  take into account the presence of both detection and nondetection in the data.
}
{We confirm that the dust mass correlates very well with the total gas mass, and find --for the first time-- that the dust mass correlates better with the atomic gas mass than with the molecular one.
We characterize
important mass ratios such as the gas fraction, the molecular-to-atomic gas mass ratio, the dust-to-total gas mass ratio (DGR), and the dust-to-stellar mass ratio, 
and study how they relate to each other, to galaxy morphology, and to gas-phase metallicity.
Only the assumption of a metallicity-dependent $X_{\rm CO}$ reproduces 
the expected decrease of the DGR with increasing morphological stage and decreasing gas-phase metallicity, with a slope of about $1$.
The DGR, the gas-phase metallicity, and the dust-to-stellar mass ratio are, for our galaxy sample, directly linked to galaxy morphology.
The molecular-to-atomic gas mass ratio 
and the DGR show a positive correlation for low molecular gas fractions, 
but for  galaxies rich in molecular gas this trend breaks  down. 
To our knowledge, this trend has never been found before, and provides new constraints for theoretical models of galaxy evolution and
a reference for high-redshift studies.
We discuss several scenarios related to this finding.
}
{The DustPedia database of late-type galaxies is an extraordinary tool for the study of the ISM scaling relations, thanks to its homogeneous collection of data for the different ISM components.  
The database is made publicly available to the whole community.}


\maketitle
 
%

\section{Introduction}
\label{sec:intro}
The global properties of nearby galaxies are related by an intricate system of correlations that form the basis of the so-called `scaling relations'. 
They allow us to study the internal physics in different galaxy populations, as well as their formation and evolutionary histories.

Among the first recognized relationships, we recall the Tully-Fisher relation for spiral galaxies \citep[][]{tully77} and the Fundamental Plane for elliptical galaxies \citep[][]{djorgovski87,jorgensen96}. 
Thanks to the flowering of spectroscopic and photometric surveys of galaxies, recent decades have seen the elucidation of new relationships among galaxy properties, such as  those
between the star formation rate (SFR) and the stellar mass in galaxies, 
the so-called main sequence (MS) of star forming galaxies
\citep[e.g.,][]{brinchmann04,daddi07,elbaz07,noeske07,santini09,santini17,peng10,rodighiero14,speagle14,whitaker14,schreiber15,tasca15,tomczak16}. Furthermore, relationships have been discovered between the stellar mass and the average oxygen abundance, the mass--metallicity (MZ) relation  
\citep[e.g.,][]{lequeux79,garnett87,Vila-Costas92,tremonti04,erb06,erb08,henry13,maier14,maier15,maier16,salim15,sanchez17}, 
and the Kennicutt-Schmidt (KS) star formation (SF) relation \citep[][]{schmidt59,schmidt63,kennicutt98a,kennicutt98b}, relating SFR and the surface density of cold gas in disks.
Both the MS and MZ have been confirmed at different redshifts, showing a change with cosmological time and thus tracing the evolution of galaxy properties with time \citep[e.g.,][]{dave11}. 

In the last few years, the number of studies dedicated to scaling relations of the interstellar medium (ISM) components has also grown 
\citep[e.g.,][]{saintonge11a,saintonge11b,catinella12,catinella13,catinella18,corbelli12,boselli14,cortese11,cortese12,santini14,cortese16,devis17a,devis17b,calette18,zuo18,cook19,ginolfi19,lin19,lisenfeld19,sorai19,Yesuf19}.
These studies have quickly become references and offer constraints for cosmological models of galaxy evolution, which are able to trace the evolution of the different gas phases \citep[e.g.,][]{gnedin09,dutton09,duttonvandenbosch09,fu10,power10,cook10,lagos11a,lagos11b,kauffmann12,bahe16,camps16,crain17,marinacci17,diemer19,stevens19}.
However, in most of the previous studies the contribution of dust to the ISM was neglected. 
The mass of the ISM is indeed made up of  $\sim$99$\%$  gas ($\sim$74$\%$ of hydrogen, $\sim$25$\%$ of helium, and $\sim$1$\%$ of heavier elements, i.e., `metals'), and  $\sim$1$\%$ 
 dust. 
Although dust occupies a small percentage in the ISM mass budget, it is a key component, driving several processes in the ISM: 
by absorbing and scattering the ionizing and nonionizing light of the interstellar radiation field, dust participates in the energetic balance that regulates the heating and cooling process 
of the ISM \citep[see e.g.,][for a comprehensive overview of the interstellar dust in nearby galaxies]{galliano18}.
The formation of molecules in the ISM happens on the surface of dust grains 
which subsequently act as a screen to protect these molecules from dissociation.
Dust also prevents the dissociation of molecules and favors gas fragmentation and the formation 
of stars \citep[e.g.,][]{scoville13}.

Dust radiates most of its energy as far-infrared (FIR) continuum emission.
The $\textit{Herschel}$ satellite has played an important role in the study of dust thanks to its superior angular resolution and/or sensitivity compared to previous FIR space and ground-based facilities 
(e.g., IRAS, ISO, $\textit{Spitzer}$, MAMBO, LABOCA);  $\textit{Herschel}$ operates right across the peak of the dust spectral energy distribution (SED, $70-500$~$\mu$m).
This makes $\textit{Herschel}$ sensitive to the diffuse cold ($T < 25$~K) dust component that dominates the dust mass in galaxies \citep[][]{devereux90,dunne01,draine07,clark15}, 
as well as to warmer ($T > 30$~K) dust radiating at shorter
wavelengths which often dominates the dust luminosity.

Making use of the $\textit{Herschel}$ data, the DustPedia project is definitively characterizing the dust properties in the Local Universe by exploiting 
a database of multi-wavelength imagery and photometry that greatly exceeds the scope 
(in terms of wavelength coverage and number of galaxies) of any similar survey \citep[][]{clark18}.   
The original DustPedia sample consists of 875 extended ($D_{25}>$1\arcmin)\footnote{$D_{25}$ is the major axis isophote at which the optical surface brightness 
falls beneath 25~mag~arcsec$^{-2}$  (we also use here $r_{25} = D_{25}/2$).} 
galaxies of all morphological types,  within $v = $~3000~km~s$^{-1}$ ($z < 0.01$), and observed by $\textit{Herschel}$  
\citep[see][for a detailed description of the DustPedia sample]{davies17}.

In the present work, we select from the DustPedia database\footnote{The DustPedia database is available at http://dustpedia.astro.noa.gr/}, 
a sample of nearby late-type galaxies to study  the correlations between the various components of the ISM.
We focus on late-type galaxies because they have -on average- a richer ISM content than early-type galaxies 
\citep[e.g.,][]{casasola04,cortese12,nersesian19}. 
In particular, more observational campaigns and single-object studies have been performed on  the molecular gas content of late-type galaxies 
and with a higher detection rate than similar investigations in early-type galaxies \citep[e.g.,][]{combes07,osullivan18,espada19}. 
Our galaxy sample covers large dynamic ranges of various galactic physical properties:
morphological stage $1 \leq$~T~$\leq 10$, $10^8$~M$_\odot$~$\lesssim$~M$_{\rm star}$~$\lesssim 10^{11}$~M$_\odot$, 
$10^{-3}$~M$_\odot$~yr$^{-1}$~$\lesssim$~SFR~$\lesssim$~60~M$_\odot$~yr$^{-1}$, and oxygen abundance $8.0 \leq$~12~+~log(O/H)~$\leq 9.5$. 
These ranges exceed those of any similar studies making our sample ideal to investigate the
ISM scaling relations.    

All the derived quantities have been collected to obtain a homogeneous and statistically significant dataset.  
In particular, we estimate all the physical properties in a common  region, namely within the optical disk of the galaxies. 
To our knowledge, this is a novel approach with respect to previous works: 
we compare different galactic properties in a very large sample of galaxies, but considering exactly the same regions within each galaxy.     
This allows us to produce consistent comparisons between masses of co-spatial galactic properties.    
In addition, DustPedia uses uniform prescriptions (i.e., models) to transform the observed quantities into dust and stellar masses (see later). 
Thus, our uniform treatment of data and modeling provides a complete and homogeneous galaxy sample able to put constraints   
on future cosmological hydro simulations predicting ISM properties and scaling relations and hopefully taking into account all ISM components. 
In this regard, we mention that an incoming DustPedia paper is focused on the comparison between DustPedia and EAGLE galaxies \citep[][]{Trvcka17}.

The paper is organized as follows.
In Sect.~\ref{sec:sample} we outline the sample selection and in Sect.~\ref{sec:data} we present data used in this work, in particular the derivation of the 
masses and their distribution as a function of morphological type.  
In Sect.~\ref{sec:scaling} we present and discuss scaling relations between the dust and atomic, molecular, and total gas masses,
and also relations between stellar and gas masses.
In Sect.~\ref{sec:ratios} we study various mass ratios,  also involving the stellar mass, as a function of different galaxy properties 
such as galaxy morphology and gas-phase metallicity. 
We highlight and summarize our main results in Sect.~\ref{sec:conclusions}.
In Appendices \ref{sec:hicodata} and \ref{sec:caveats}, we collect the main properties of gas data and describe caveats and uncertainties associated with our analysis.

\section{The galaxy sample}
\label{sec:sample}
Our selection is based on the following criteria applied to the DustPedia database:
$i)$ Hubble stage T ranging from 1 to 10;  
$ii)$ available global flux at 250~$\mu$m $>$3$\sigma$;
$iii)$ no significant contamination to the global flux from 
nearby galactic or extragalactic sources, and no images with 
artifacts or insufficient sky coverage for a proper estimate of the target/sky levels.
The flux selection, required by the companion work of \citet{bianchi19},
cuts about 20\% of all late-type DustPedia galaxies, while the quality selection cuts only 5\%. 
We further discuss the effect of the flux selection in Sect.~\ref{sec:selection}.
The selected sample is composed of 436 DustPedia late-type galaxies. 

Our sample is divided into bins following the Hubble stage T. 
The morphology indicator  T is obtained from the HyperLEDA database \citep[][]{makarov14}
and it can be a noninteger since for most objects  the final T is averaged over various estimates available in the literature. 
Following \citet{bianchi18}, we use all objects in the range $\left [{\rm T} - 0.5, {\rm T} + 0.5\right)$ to define a sample characterized by a given integer T (e.g., the Sa sample defined by T~=~1 include all objects with ${\rm 0.5 \leq T < 1.5}$).  
We use distances and other galaxy properties (e.g., $r_{25}^{1}$) collected by \citet{clark18} and distributed together with the DustPedia photometry.
  
About 50$\%$ of sample galaxies are classified as interacting systems.
This definition of interacting galaxies is very broad, including pair and group members and parents of a companion galaxy according to the NED homogenized classification.
According to the NED classification, approximately $12$\% of sample galaxies are low-luminosity active galactic nuclei (AGNs, e.g., L$_{\rm X} < 10^{42}$~erg~s$^{-1}$), including Seyferts and LINERs, and
$\sim$2$\%$ are starbursts.
Table~\ref{tab:sample} collects the number of galaxies in the main sample for each morphological type,  with some interaction or activity of our sample galaxies. 

As shown in the following section, most of the galaxies in the main sample have high-quality \textit{Herschel} data (dust content). In addition, most of them have been observed at 21-cm, and therefore we know their total atomic gas content, and  an estimate of their stellar mass is available.
Moreover, for an important fraction of them,  gas-phase metallicities and $^{12}$CO emission line data (and therefore molecular gas mass content) are available.

\begin{table}
\caption{\label{tab:sample} 
{
Classification of the main properties of the DustPedia late-type galaxy sample.
}}
\centering
\begin{tabular}{llccc}
\hline\hline
Type                                    & No. galaxies  \\
\hline
All                                             & 436  \\
Sa (T = 1)                              & 48  \\
Sab (T = 2)                             & 37 \\
Sb (T = 3)                              & 59 \\
Sbc (T = 4)                             & 56 \\
Sc (T = 5)                                      & 62 \\
Scd (T = 6)                             & 70 \\
Sd (T = 7)                                      & 37 \\ 
Sdm (T = 8)                             & 26 \\ 
Sm (T = 9)                              & 19  \\ 
Irr (T = 10)                            & 22   \\       
Interacting $^{(a)}$                    & 218 \\
Low-luminosity AGN $^{(b)}$     & 52 \\
Starbursts $^{(c)}$                     & 10 \\
\hline
\hline
\end{tabular}
\tablefoot{
$^{(a)}$~See Sect.~\ref{sec:sample} for the definition of interacting galaxy.
$^{(b)}$~Including Seyferts and LINERs.
The sub-classes of galaxies collected in $^{(a)}$, $^{(b)}$, and $^{(c)}$ are defined according to NED classifications.
}
\end{table}

\section{The data}
\label{sec:data}
For the present work, we collect from the literature observations of molecular (Sect.~\ref{sec:co}) 
and atomic gas (Sect.~\ref{sec:hi}). 
We describe here the data homogenization process and the estimate of the gas masses within $r_{25}$. 
We use this aperture since it contains most of the dust (and stellar) luminosity
\citep[e.g.,][]{pohlen10,casasola17,clark18}. 
We also use dust and stellar masses, and gas-phase metallicities as described in a companion work of the DustPedia collaboration (Sect.~\ref{sec:dustmass}). 

We stress that the dust and stellar fluxes of \citet{clark18} used to derive the corresponding masses of DustPedia galaxies
can refer to radii beyond the optical disk. 
The aperture-fitting process of \citet{clark18} is indeed based on an elliptical aperture in every band for a given target, and these apertures are then
combined  to provide a final aperture for the target.
This approach has provided consistent optical coverage for all DustPedia galaxies \citep[see][for more details]{clark18}, and the study focusing on DustPedia face-on spiral  galaxies by \citet{casasola17} showed that the total dust and stellar masses are a 
good approximation of their values within the optical radius. Table~\ref{tab:tot-data} summarizes the data available for our galaxy sample.

\begin{table}
\caption{\label{tab:tot-data} 
Total data collected for the studied galaxy sample.}
\centering
\begin{tabular}{llcccc}
\hline\hline
Data    & No. galaxies $^{(1)}$ \\
\hline
250~$\mu$m                                      & 436 \\
$^{12}$CO $^{(2)}$                              & 255 (210)  \\
$^{12}$CO(1--0) $^{(3)}$                        & 243 (200)  \\
$^{12}$CO(2--1) $^{(3)}$                        & 5 (3)  \\
$^{12}$CO(3--2) $^{(3)}$                        & 7 (7)  \\
H{\sc i}                                                & 433  (422)$^{(4)}$  \\
$^{12}$CO $\&$ H{\sc i}                 & 255 \\
M$_{\rm d}$                             & 432 \\
M$_{\rm d}$, $^{12}$CO $\&$ H{\sc i} detections & 202$^{(5)}$ \\
M$_{\rm star}$                                  & 432 \\
12~+~log(O/H)$_{\rm N2}$$^{(6)}$        & 339 \\
\hline
\hline
\end{tabular}
\tablefoot{
$^{(1)}$~Number of galaxies of the sample with a given type of data. 
For $^{12}$CO and H{\sc i} data the number of detections is given between brackets.
$^{(2)}$~Information including the transitions $^{12}$CO(1--0), $^{12}$CO(2--1), and $^{12}$CO(3--2). 
$^{(3)}$~Information for each $^{12}$CO transition. 
$^{(4)}$~Uncertainties on H{\sc i} data for 28 out of 422  detections are not available.
$^{(5)}$~These data define the gaseous sample (see Sect.~\ref{sec:hi}). 
$^{(6)}$~Oxygen abundances from the empirical calibration N2 of \citet{pettini04} (see Sect.~\ref{sec:metallicity}).
}
\end{table}

\subsection{The molecular gas}   
\label{sec:co}
\label{sec:gasdata}
\label{sec:masses}

For the molecular gas, we bring together $^{12}$CO observations from a wide variety of sources (see Table~\ref{tab:refs}).
We use data on  $^{12}$CO(1--0), $^{12}$CO(2--1), 
and $^{12}$CO(3--2) emission lines, observed at 2.6~mm (115~GHz), 1.3~mm (230~GHz), and 
0.87~mm (345~GHz), respectively.  For each galaxy, our first choice is the $^{12}$CO(1--0) emission line, when available and of high quality (e.g., high S/N ratio). The second choice is the $^{12}$CO(2--1) line, and the third and final choice is the $^{12}$CO(3--2) line.
We report $^{12}$CO(2--1) and $^{12}$CO(3--2) emission lines to the $^{12}$CO(1--0) one
by adopting the $^{12}$CO line ratios $R_{21}$ ($=I_{21}/I_{10}$), 
$R_{32}$ ($=I_{32}/I_{21}$), and $R_{31}$ ($=I_{32}/I_{10}$). When available, we use values measured for a given galaxy
\citep[e.g., NGC~1808,][]{aalto94}. Otherwise, we assume $R_{21} = 0.7$, as determined in the HERACLES survey 
(Leroy et al. 2009, see also Casasola et al. 2015)
and extensively used in studies of nearby late-type galaxies \citep[e.g.,][]{schruba11,casasola17}; and $R_{32} = 0.36$ and $R_{31} = 0.18$, determined by \citet{wilson12}.     

In most cases, observations are given in terms of different temperature scales (see Table~\ref{tab:CO-telescopes}) which can be transformed 
into a common scale following \citet{boselli14} and the prescriptions of \citet{kutner81}.
We homogenize the dataset by transforming all temperatures into $^{12}$CO(1--0) fluxes, with $S_{\rm{CO}}$ in units of Jy~km~s$^{-1}$, following Table~\ref{tab:CO-telescopes}.
In two cases (NGC~3198 and NGC~3256), we extracted the H$_2$ mass (M$_{{\text{H}}_2}$) from the literature. 
These masses were properly transformed  into the total $^{12}$CO flux,    
by adopting the distance and the CO-to-H$_2$ conversion factor used in the original reference.

\subsubsection{Derivation of the $^{12}$CO flux within the optical disk}
\label{sec:derco}
Most of our galaxies ($\sim$92$\%$) have single-beam $^{12}$CO observations 
usually pointed on the center of the  galaxy and  with a beam smaller than $r_{25}$. 
The $^{12}$CO flux must then be corrected for aperture effects to derive the total line flux.
Since our sample contains highly inclined galaxies ($i > 60^{\circ}$), we follow \citet{boselli14} and
assume that the $^{12}$CO emission is well described by an exponential decline both along the radius and above the galactic plane: 
\begin{eqnarray}
{S_{\rm{CO}}(r,z)} &=& S_{\rm{CO}}(0){\rm e}^{-r/r_{\rm CO}}{\rm e}^{-|z|/z_{\rm CO}},
\label{eq:SCOrz}
\end{eqnarray}
with $r_{\rm CO}$ and $z_{\rm CO}$ being the scale-length and scale-height of the disk, respectively. 
For galaxies with low inclination, the method is analogous to the standard 2D approach, such as that developed by \citet{lisenfeld11}. 
The method is  based on the fact that the $^{12}$CO emission of nearby mapped galaxies can be
described by an exponential disk: its radial scale-length $r_{\rm CO}$ correlates well with the optical scale-length 
of the stellar disk and with $r_{25}$ \citep{lisenfeld11,casasola17}. 
By assuming an average ratio $r_{\rm CO}$/$r_{25}$, the $^{12}$CO 2D distribution can be simulated, 
convolved with the beam profile in order to scale it to the observation in the center of the galaxy, 
and finally integrated to obtain the total $^{12}$CO flux \citep[e.g.,][]{stark86,young95,corbelli12}. 
We assume here  $r_{\rm CO}/r_{\rm 25} = 0.17 \pm 0.03$, as derived by \citet{casasola17} 
for a sub-sample of DustPedia face-on spiral galaxies and  consistent with the values of \citet{leroy08} and  \citet{lisenfeld11}.
For galaxies closer to the edge-on case, the thickness of the disk must be taken into account: 
we use $z_{\rm CO}/r_{\rm 25} = 1/100$, defined by \citet{boselli14} from CO and dust emission models and observations of edge-on galaxies.

A few galaxies ($\sim$4$\%$ of the sample) instead have angular sizes smaller than the  $^{12}$CO beam. 
When not clearly specified in the original reference, we have corrected these $^{12}$CO data for beam dilution 
$T_{\rm B} = T_{\rm mb} \left(\frac{\theta_{\rm S}^2 + \theta_{\rm beam}^2}{\theta_{\rm S}^2} \right)$, where $T_{\rm B}$ is the brightness temperature and $\theta_{\rm S}$ the source size.

\subsubsection{CO-to-H$_2$ conversion and mass of H$_2$}
The mass of molecular hydrogen M$_{{\text{H}}_2}$ is derived under the assumption of optically thick $^{12}$CO(1--0) emission, through the formula:
\begin{eqnarray}
{\rm M_{{\text{H}}_2}} &=& 3.9 \times 10^{-17} \times X_{\rm CO} \times S_{\rm{CO}} \times D^2,
\label{mh2}
\end{eqnarray}
\noindent
where M$_{{\text{H}}_2}$ is in units of M$_\odot$, $S_{\rm{CO}}$ is the $^{12}$CO(1--0) flux in units of Jy~km~s$^{-1}$, aperture-corrected as in Sect.~\ref{sec:derco}, and $D$ is the galaxy distance extracted from the DustPedia database in units of megaparsecs.   
The derivation of the total molecular mass in Eq.~(\ref{mh2}) requires knowledge of the CO-to-H$_2$ conversion factor, $X_{\rm CO}$. 
We make two different assumptions on $X_{\rm CO}$: 
{\em i)} a constant value in agreement with the recommended value for Milky Way-like disks, which represents most galaxies of our sample,  $X_{\rm CO} = 2.0 \times 10^{20}$ cm$^{-2}$ (K~km~s$^{-1}$)$^{-1}$  with $\pm30\%$ uncertainty from \citet{bolatto13};
 and {\em ii)} a metallicity-dependent $X_{\rm CO}$ as in \citet{amorin16}. 
The study of these two assumptions on $X_{\rm CO}$ provides a conservative range of molecular gas mass estimations 
that demonstrates how uncertain the molecular gas mass derivation is. 

As derived  by  observations  \citep[e.g.,][]{bolatto08,bolatto13,casasola07,magrini11,accurso17,remy17} and by theoretical models 
\citep[e.g.,][]{glover11,narayanan12,gong18}, the $X_{\rm CO}$ conversion factor can vary due to effects of metallicity, gas temperature and abundance, 
optical depth, cloud structure, cosmic ray density, and ultraviolet radiation field.
The dependence of $X_{\rm CO}$ on the abundance of the heavy elements is particularly debated in the literature: there are several relationships of $X_{\rm CO}$ with  metallicity, showing a range of behaviors 
\citep[e.g.,][]{wilson95,arimoto96,barone00,israel00,boselli02,magrini11,schruba12,hunt15b,amorin16}. 
These relationships, although different, show a general increase of $X_{\rm CO}$ with decreasing metallicity.
Particular attention is paid to galaxies with metallicity below $\sim$20$\%$ solar, such as blue compact dwarfs, for which very deep observational campaigns are needed to detect the $^{12}$CO(1--0) line \citep{madden13,cormier14,hunt14,hunt15b,hunt17}. 
Since the abundances in our sample are not particularly  extreme (see Sect.~\ref{sec:metallicity}), we adopted the metallicity-dependent CO-to-H$_2$ conversion factor by \citet{amorin16}, which is derived combining low-metallicity starburst galaxies with more metal-rich galaxy objects, including the Milky Way and Local Volume galaxies from \citet{leroy11}: $X_{\rm CO} \propto (Z/Z_\odot)^{-1.5}$ 
(see the fit in Fig.~11 of \citealt{amorin16}, where the conversion factor is given in terms of the equivalent $\alpha_{\rm CO} [M_\odot \; \mathrm{pc^{-2} \;(K \;km \;s^{-1})^{-1}}] = 1.6 \times 10^{-20} 
\; X_{\rm CO} [\mathrm{cm^{-2} \;(K \;km \;s^{-1})^{-1}}]$). The power law of the calibration of \citet{amorin16} is also in qualitative agreement with previous determinations \citep[e.g.,][]{genzel12,schruba12} 
and model predictions \citep[e.g.,][as presented in Sandstrom et al. 2013]{wolfire10}.    

Uncertainties in M$_{{\text{H}}_2}$ are calculated as the quadrature sum of the uncertainty on the $^{12}$CO(1--0) flux and on the $X_{\rm CO}$ conversion factor. 
Under the assumption of a metallicity-dependent $X_{\rm CO}$, uncertainties in M$_{{\text{H}}_2}$  also take into account uncertainties on the metallicity and on the 
calibration of $X_{\rm CO}$ with the metallicity.

\subsection{The atomic gas}   
\label{sec:hi}
\citet{devis19} collected all available H{\sc i} 21~cm emission line observations from the literature 
for the whole DustPedia sample. 
The H{\sc i} fluxes $S_{\rm{HI}}$ are provided in units of Jy~km~s$^{-1}$; as 
for $^{12}$CO data, they come from various telescopes characterized by different beams. 
Table \ref{tab:HI-telescopes} collects telescopes and beams of the H{\sc i} data used in the current work.
The mass of the atomic gas, M$_{\rm{HI}}$, under the assumption of optically thin H{\sc i} emission,  is given by
\begin{eqnarray}
{\rm M_{\rm{HI}}} &=& 2.356 \times 10^5 \times S_{\rm{HI}}\times D^2,
\label{mhi}
\end{eqnarray}
\noindent
where M$_{\rm{HI}}$ is in units of M$_\odot$ and the distance $D$ is in megaparsecs.
Uncertainties in M$_{\rm{HI}}$ have been calculated from the uncertainty on the H{\sc i} flux. 

In highly inclined galaxies, H{\sc i} emission might not be optically thin. 
\citet{haynes84}  derived empirical corrections for galaxies of different axial ratios $b/a$ and found that the 
H{\sc i} flux could be underestimated by more than 20\% for galaxies with $b/a\le 0.25$ and type Sc-Sd only.
Our main sample contains only 12 such galaxies, a number that cannot alter the results of this study. 
We also find no significant overestimation of the dust-to-gas mass ratio for galaxies seen at edge-on inclinations
(we have used the values derived by \citealt{mosenkov19})  if we use either H{\sc i} only, or the total gas mass
including the molecular phase. 
We therefore conclude that the assumption of the optically thin H{\sc i} 
is valid and the H{\sc i} masses used in this work are not severely underestimated.

Contrary to $^{12}$CO, H{\sc i} observations typically cover a region similar to or larger than $r_{25}$ (compare the beam sizes in Tables~\ref{tab:CO-telescopes} 
and \ref{tab:HI-telescopes}). 
In particular, 77\% 
of the available H{\sc i} observations refer to larger sky areas. 
Therefore, before estimating the mass, it has been necessary to correct (i.e.,\ reduce)  $S_{\rm{HI}}$ to 
obtain the H{\sc i} line flux within the optical radius.

We estimate the  H{\sc i} flux within $r_{25}$  adopting the model of the radial H{\sc i} 
surface density profiles of \citet{wang14} obtained from azimuthally averaged radial profiles of H{\sc i} gas in 42 galaxies from the Bluedisk sample \citep{wang13}. 
Their model for the H{\sc i} profiles  is an exponential function 
of radius in the outer regions with a depression towards the center, and it scales with R1\footnote{R1 is 
the radius where the H{\sc i} surface density ($\Sigma_{\rm HI}$) is equal to 1~M$_\odot$~pc$^{-2}$.}.
In the outer regions, the radial H{\sc i} surface density profiles are highly homogeneous for all galaxies, and exponentially declining parts have a scale-length 
of $\sim$0.18~R1 \citep[see Fig.~10 in][]{wang14}.
Similar results were found by \citet{cayatte94} and \citet{martinsson13}, for example. 
Following \citet[see in particular their Eqs.~9 and 10]{wang14} and by integrating $\Sigma_{\rm HI}(r)$ up to $r_{25}$/R1,
we derived the H{\sc i} mass within $r_{25}$ 
for our galaxies with observations of larger extent. 
For these galaxies, the H{\sc i} mass reduces by about $\sim$30--35$\%$.
The adopted method is preferred to other available methods \citep[see, e.g.,][]{bigiel12} since it is 
applicable to almost the totality of our galaxy sample, it is based only on H{\sc i} observations, and it is valid for both interacting and noninteracting galaxies.  

It is important to stress that by focusing within the optical radius of late-type galaxies, the H{\sc i} mass suffers the strongest reduction compared to the other ISM components and galaxy properties.  
It is also well known that the atomic gas beyond $r_{25}$ is very interesting.   
The need to derive H{\sc i} mass within the common radius of $r_{25}$ is also dictated by the fact the original H{\sc i} fluxes 
cover regions not only typically larger than $r_{25}$ but also larger than $r_{25}$ in a different way. 
Therefore, the choice of adopting the optical radius as size to explore scaling relations represents a solid normalization parameter.
As already mentioned in Sect.~\ref{sec:intro}, we are looking at different properties in a very large sample of galaxies, and always 
in exactly the same region of each galaxy.
This is a novel approach with respect to those adopted in well-known surveys such as COLD GASS and xGASS 
\citep[e.g,][]{saintonge16,catinella18}, where no restrictions were applied to fluxes used to derive the corresponding masses.      

As typically done in similar studies, we assume that the total gas mass is the sum of the atomic and molecular gas masses, 
corrected for the helium contribution by multiplying by a factor of 1.36
(M$_{\rm tot\,gas}$ = 1.36~$\times$~[M$_{{\rm{H}}_2}$ + M$_{\rm{HI}}$]).

We refer to the subsample of galaxies that have detected molecular and atomic gas masses, as well as dust masses, as the gaseous sample; there are 202 galaxies in this sample (see Table~\ref{tab:tot-data}).

\subsection{Other DustPedia quantities}
\label{sec:dustmass}
\label{sec:metallicity}
Dust masses (M$_{\rm d}$) and stellar masses (M$_{\rm star}$) of DustPedia galaxies
were obtained through the modeling of their spectral energy distribution (SED). 
The Code Investigating GALaxy Evolution (CIGALE)\footnote{Version 0.12.1, available at https://cigale.lam.fr.} was used \citep[][]{boquien19}.
CIGALE allows us to model the SED of a galaxy by choosing a variety of modules for the stellar, gas, and dust emission, and for dust attenuation \citep[][]{noll09,roehlly14,boquien19}.
In the version of CIGALE adopted by the DustPedia collaboration, the dust emission templates are computed from the DustPedia reference grain model \citep[THEMIS\footnote{The 
Heterogeneous Evolution Model for Interstellar Solids, http://www.ias.u-psud.fr/themis/index.html};][]{jones17}.
For the full description of the DustPedia/CIGALE sample selection and modeling we refer to Nersesian et al. (2019; see also Bianchi et al. 2018).

The estimates and uncertainties of M$_{\rm d}$ and M$_{\rm star}$ from CIGALE are based on Bayesian statistics.
Four galaxies of our sample (NGC~0253, NGC~4266, NGC~4594, NGC~7213) lack dust and stellar masses (see Table~\ref{tab:tot-data}); they have not been fitted by CIGALE
because their optical and mid-infrared (MIR) fluxes are contaminated or because of null coverage in the MIR and FIR bands.

Global oxygen abundances are available for a significant fraction of DustPedia galaxies \citep{devis19}.
As in most studies, we assume that the oxygen abundance is a good tracer of the total gas-phase metallicity.
This assumption is justified by the fact that oxygen is the most abundant element besides H and He in the Universe. 
In addition, it is among the dominant constituents of dust \citep[e.g.,][]{savage96,jenkins09}. 
We adopt a solar oxygen abundance of 12 + log(O/H) = $8.69 \pm 0.05$ from \citet{asplund09}.   
The global metallicities provided by \citet{devis19} are given at $r = 0.4 \times r_{25}$ due to the statistically confirmed relationship between the luminosity-weighted integrated metallicity and 
the characteristic abundance \citep[][]{kobulnicky99,pilyugin04,moustakas06,moustakas10}.   

Among the available metallicity determinations in \citet{devis19}, 
we select the empirical calibration N2 from \citet{pettini04}, based  on the ratio of [\ion{N}{ii}] $\lambda$~6584~\AA\, and H$\alpha$, since it is available for most galaxies of our sample 
(339/436 galaxies, see Table~\ref{tab:tot-data}).  
The N2 global metallicities of our galaxy sample span from 12+log(O/H)$_{\rm N2}$~$=$ 8.0 to 9.5.

\subsection{The morphological-type dependence of the mass distributions}
\label{sec:selection}

\begin{figure}
\resizebox{\hsize}{!}{
\includegraphics[trim=30 10 220 10,clip]{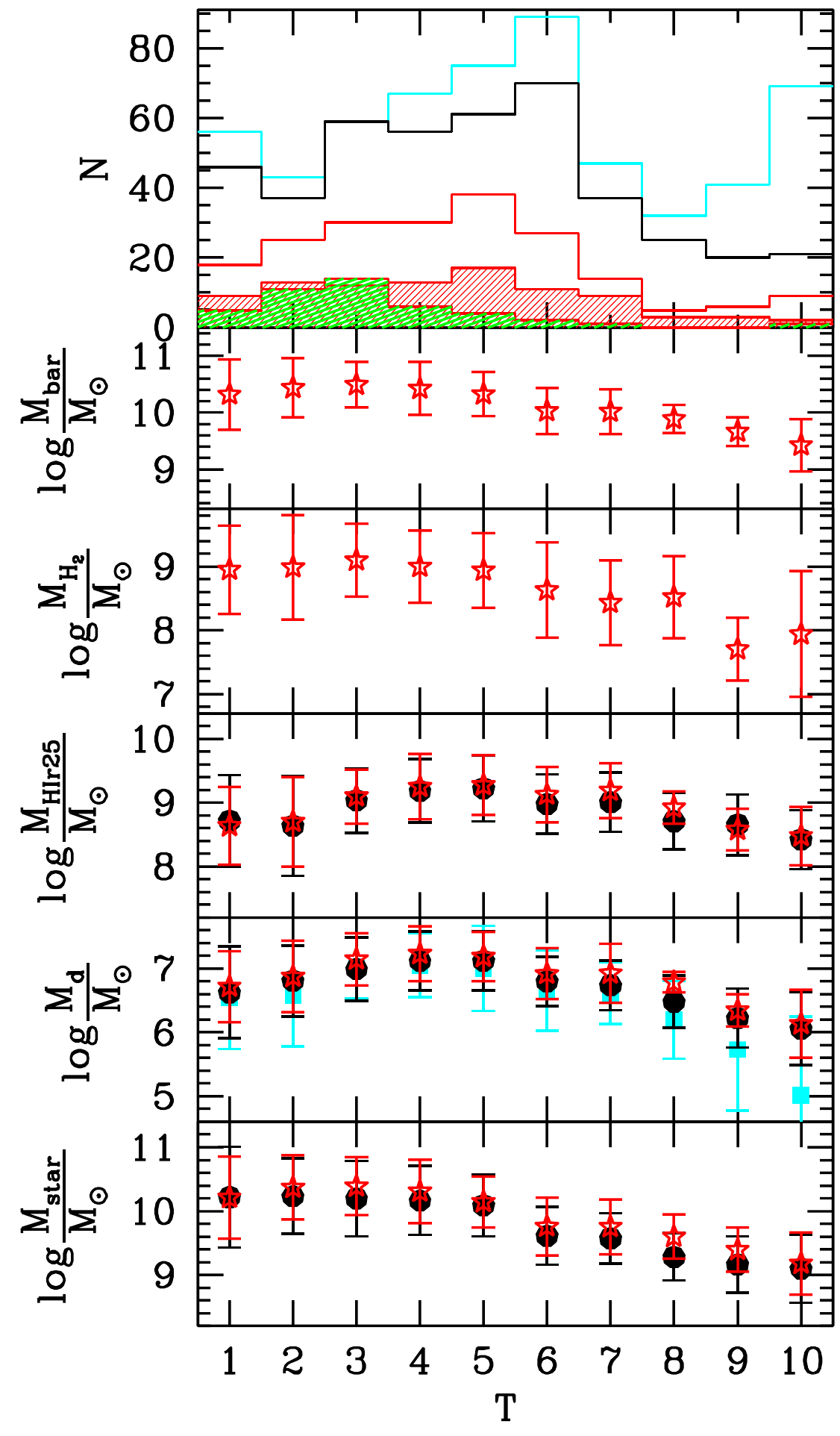}
}
\caption{Number of galaxies and the mean of the log of baryonic, H$_2$, H{\sc i}, dust, and 
stellar masses as a function of morphological stage 
in our main sample (black lines and filled dots) and in the gaseous sample (red lines and stars symbols). 
The H{\sc i} gas mass shown is inside $r_{25}$. 
We have not corrected gas masses for helium except for the baryonic masses which include the atomic gas beyond $r_{25}$. 
The shaded red and green areas in the upper panel point out the distribution of the number of interacting (92) and active (44 AGN or starbursts) 
galaxies, respectively, in the gaseous sample.
We also show the number of objects and the dust mass distribution for all DustPedia galaxies (cyan lines and filled squares).}
\label{fig:distri}
\end{figure}

\begin{figure}[ht]
\centering
\includegraphics[width=0.50\textwidth]{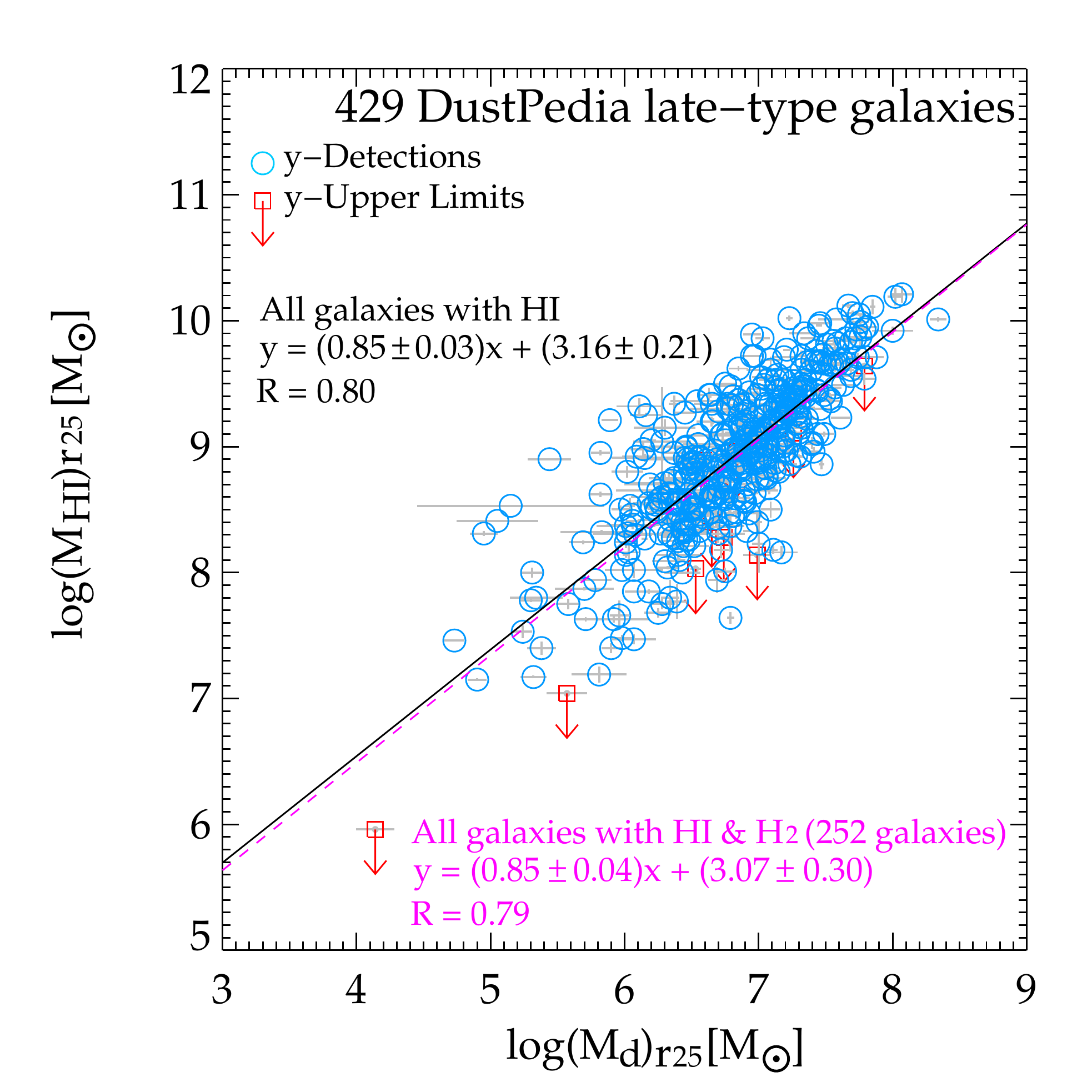}
\caption{
Scaling relation between dust and H{\sc i} masses within $r_{25}$, in logarithmic scale, for 429 DustPedia late-type galaxies
(those with dust and H{\sc i} mass data). 
The light blue circles are log(M$_{\rm HI}$) detections with error bars drawn in light gray. 
Red squares with downward arrows show log(M$_{\rm HI}$) upper limits. 
The (black) continuum line is the line fit derived for the plotted 429 galaxies, and the (magenta) dashed line that for
the sample galaxies with dust, H{\sc i}, and H$_2$ data (252 galaxies, those plotted in both panels of Fig.~\ref{fig:logmd-logmh2totgas}).
Equations of the line fits and the Pearson correlation coefficients $R$ are also given in the figure.
}
\label{fig:logmd-logmhi}
\end{figure}

Having defined the sample and the masses relative to each sample galaxy we now show in Fig.~\ref{fig:distri} the number of galaxies, and the mean value of the 
baryonic, molecular, and atomic hydrogen, dust, and stellar mass as a function of morphological type. 
Molecular gas mass is derived under the assumption of constant $X_{\rm CO}$.
Atomic hydrogen mass is shown within $r_{25}$ while baryonic mass includes stars, and atomic and molecular gas masses (with helium) throughout the galaxy, 
beyond $r_{25}$.
The number of galaxies in each morphological class is shown for the whole sample and for the gaseous sample.
In addition, we also show the distribution of active and interacting galaxies. 
The atomic hydrogen gas, dust, and stellar mass distribution are computed for the main and the gaseous sample (in black and red, respectively), 
while the molecular hydrogen gas and baryonic mass distribution are shown only for the gaseous sample.
The distribution of the data according to galaxy morphology shows that:
\begin{itemize}
\item The gaseous sample is representative of the main sample.
\item Dust and atomic gas masses peak for Sbc--Sc-type (T~=~4--5) galaxies and decline for earlier and later types, while the stellar mass drops more steadily with T. 
This confirms the results of R{\'e}my-Ruyer~et~al.~(2014, see their Appendix) and \citet{nersesian19}.
\item Molecular gas masses follow the distribution of stellar mass, decreasing for high T and  underlining the role of stars in enhancing the formation of molecules by compressing the ISM \citep{blitz04,wong13}.
\item The baryonic mass is dominated by stars for earlier-type galaxies, while both gas and stars contribute for later types. 
The mean value of the baryonic mass decreases by a factor of about ten going from the peak value of $3\times 10^{10}$~M$_\odot$ for Sb galaxies to the irregular galaxies. 
\end{itemize}

We reiterate that galaxies in the main sample are required to be 
detected at 250 $\mu$m. Thus, our main (and gaseous) sample 
might be biased against objects with smaller dust masses. Thanks to SED
fitting and to the wide wavelength coverage, dust masses 
are available for most of the DustPedia sample, even for objects 
not detected  or observed at 250 $\mu$m (they are missing for 
only 5\% of DustPedia late-type galaxies because of flux-quality requirements; Sect.~\ref{sec:data}). Even though our main sample includes only 75\% 
of all DustPedia late-type galaxies, the distribution of dust mass versus
morphology is compatible with that for the full sample (cyan datapoints 
in Fig.~\ref{fig:distri}), for all but the later types
(which suffer from the flux limit to a greater extent: only 45\% of DustPedia 
galaxies with $T\ge8$ survive the cut). 
However, we do not expect this bias to significantly affect our 
results due to the large scatter in the data and to equivalent 
flux-dependent biases which will not greatly affect the dust-to-star 
and dust-to-gas ratios
(e.g., the 250~$\mu$m flux selection leads to a
similar overestimation of the stellar mass of later-type galaxies; not shown).

\section{The ISM scaling relations in the Local Universe}
\label{sec:scaling}
In this section, we present and discuss different scaling relations between the masses
of all ISM components (dust and gas)
in combination with the gas-phase metallicities and stellar masses.
Even though it is not possible to separate the dust masses associated with each gas phase, 
here we study the correlation of the total dust mass with atomic and molecular gas separately in order to highlight trends in the ranges
where either of the gas components dominates.
As explained in the previous sections, all data are uniformly homogenized and all masses refer to the values within $r_{25}$.

\subsection{Scaling relations between the masses of gas and dust}
\label{sec:scal-mass}
\label{sec:survival}
\begin{figure*}
\centering
\includegraphics[width=0.49\textwidth]{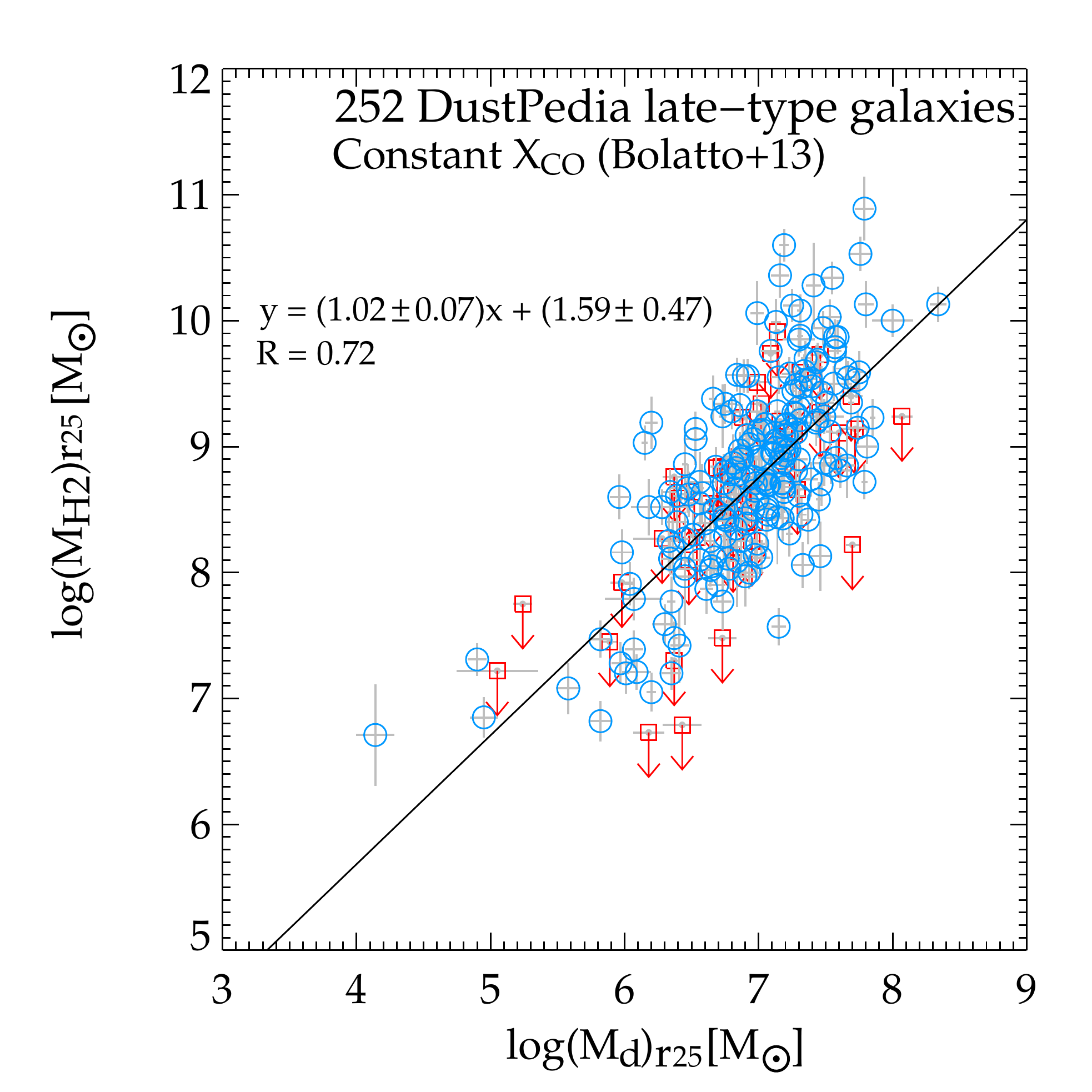}
\includegraphics[width=0.49\textwidth]{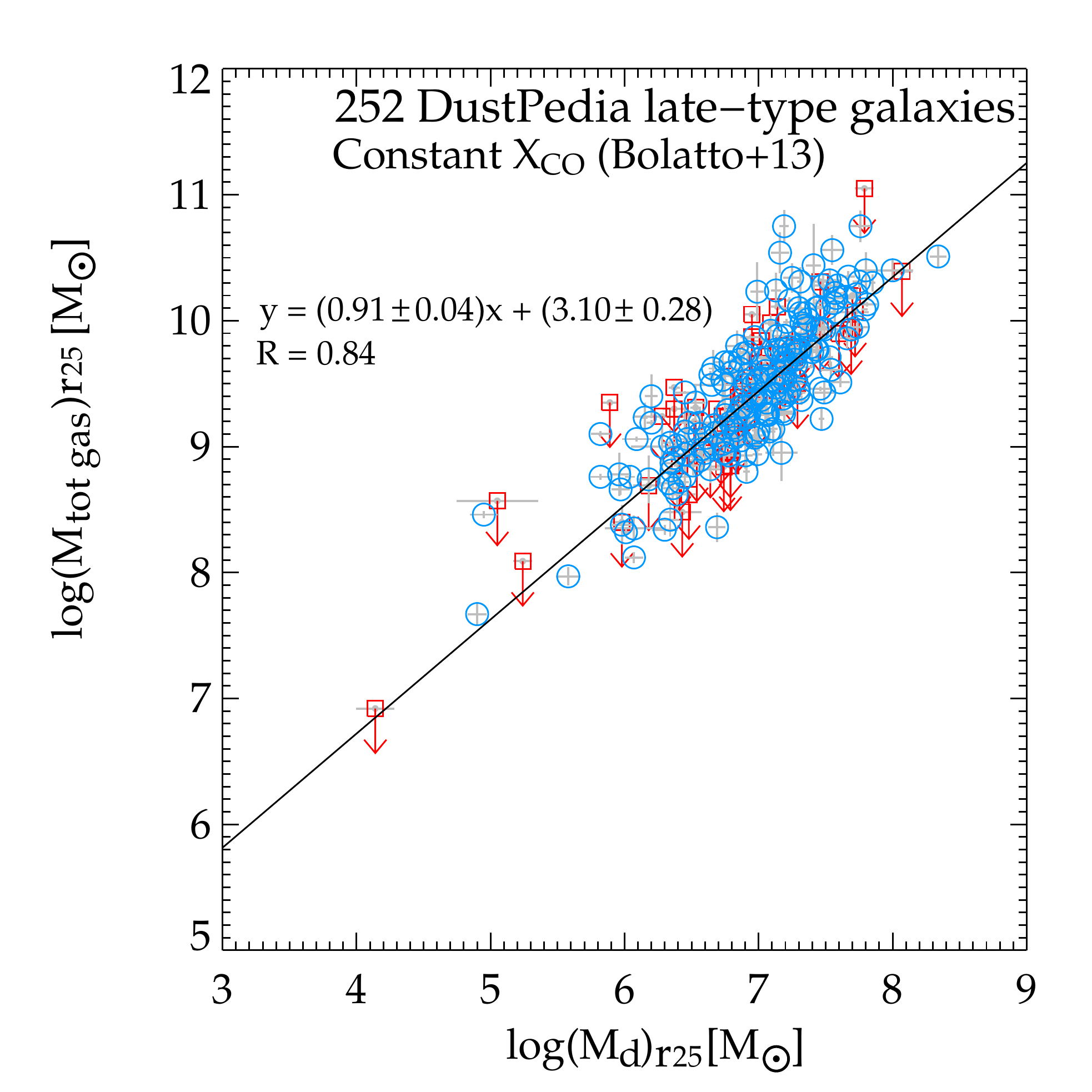}
\caption{
\textit{Left panel:}
Same as Fig.~\ref{fig:logmd-logmhi} for the scaling relation between dust and H$_2$ masses
for 252 DustPedia late-type galaxies (those with dust and H$_2$ data).
The H$_2$ gas masses were derived assuming the constant $X_{\rm CO}$ \citep[value from][]{bolatto13}. 
\textit{Right panel:} Same as \textit{Left panel} but for the scaling relation between dust and total gas masses.
}
\label{fig:logmd-logmh2totgas}
\end{figure*}

We study the ISM scaling relations between the dust mass and atomic, molecular, and total gas mass
under two assumptions on the $X_{\rm CO}$ conversion factor, that is, by adopting 
a constant $X_{\rm CO}$ and a metallicity-dependent one (see Sect.~\ref{sec:masses}).
In the following plots (Figs.~\ref{fig:logmd-logmhi} and \ref{fig:logmd-logmh2totgas}),  we fit the data in the logarithmic space:
\begin{eqnarray}
{\rm log}({\rm M}{\rm_{gas}})_{r25} = q + m\,\times\, {\rm log}({\rm M}_{\rm d})_{r25},
\label{eq:logm}
\end{eqnarray}
\noindent
where $m$ and $q$ are the slope and the intercept coefficients of the fit, respectively, and 
M$_{\rm gas}$ is the mass of atomic, molecular, or total gas in units of M$_\odot$.  
The coefficients of the fits, the Pearson correlation coefficients $R$, and the number of sample galaxies taken into account for each correlation 
are given in the figures and in Table~\ref{tab:coeff}. 
Based on the $t-$test used to establish if the coefficient $R$ is significantly different from zero, we stress that 
the significance $P$ of all coefficients $R$ provided in this work are $P < 0.0001$ (i.e., very highly significantly different from zero). 
All fits in this section are obtained using the `survival analysis' \citep[][]{feigelson12}, which consists in an ensemble of statistical methods that take into 
account the presence of `censored' data points (in our case, upper limits for $\sim$18\% and $\sim$2\% of the total $^{12}$CO and H{\sc i} masses, respectively). 
We perform the survival analysis using the EM algorithm, available through 
the Astronomy Survival Analysis (ASURV) package \citep[][]{Feigelson85,Isobe86} in the IRAF\footnote{IRAF (Image Reduction and Analysis Facility) is distributed 
by the National Optical Astronomy Observatory, which is operated by the Association of Universities for Research 
in Astronomy (AURA) under a cooperative agreement with the National Science Foundation.} environment
\citep[see][for a recent use of the survival analysis in the context of galaxy scaling relations]{Yesuf19}.

Figure~\ref{fig:logmd-logmhi} shows that dust and H{\sc i} masses are well correlated within the optical disk ($R = 0.79 - 0. 80$). 
The slope of this relationship is sublinear ($m = 0.85 \pm 0.03$, for 429 galaxies) and it remains unchanged if we consider the smaller sample of galaxies 
with both H{\sc i} and H$_2$ data ($m = 0.85 \pm 0.04$, for 252 galaxies).   
This latter result excludes the presence of possible biases due to a larger sample of galaxies with available H{\sc i} data with respect to the CO one (see Table~\ref{tab:tot-data}).
The nonlinear relation between dust and H{\sc i} masses is driven by a few galaxies with low dust-to-H{\sc i} mass ratio for  M$_{\rm HI}< 3\times 10^{9}$~M$_\odot$.
The left panel of Fig.~\ref{fig:logmd-logmh2totgas} shows that dust and H$_2$ masses, under the assumption of the constant $X_{\rm CO}$, 
are well correlated with a linear slope
($m = 1.02 \pm 0.07$, $R = 0.72$, for 252 galaxies).
The significance of the difference between the coefficients $R$ of the dust-H{\sc i} and dust-H$_2$ correlations  is $P-$value = 0.0168.
Since this $P-$value is less than 0.05 (significance level of 0.05), 
we conclude that the two $R$ differ significantly.
This allows us to infer that dust and H{\sc i} are better correlated than dust and H$_2$.   
The \textit{Right panel} of Fig.~\ref{fig:logmd-logmh2totgas} shows the relationship between dust and total gas masses: a slope of $m = 0.91 \pm 0.04$ ($R = 0.84$, for 252 galaxies).  
The significance of the difference between the coefficients $R$ of the dust-H{\sc i} and dust-total gas correlations is 
$P-$value = $0.0949 - 0.1236$ (higher than 0.05), meaning that the dust is correlated with H{\sc i} and total gas 
with similar statistical significance.       
On the contrary, the significance of the difference between the coefficients $R$ of the dust-H$_2$ and dust-total gas 
correlations is $P-$value = 0.0005, implying that dust and total gas are better correlated than dust and H$_2$. 
We performed a similar analysis with a metallicity-dependent $X_{\rm CO}$, finding similar 
relationships although with lower $R$, and again a slightly better correlation between dust and total gas masses. 
The results of the fits and correlation coefficients are collected in Table~\ref{tab:coeff}.

\begin{table*}
\caption{\label{tab:coeff} 
Main properties of the fitting lines of the scaling relations presented in Sect.~\ref{sec:scal-mass}.}
\centering
\begin{adjustbox}{max width=\textwidth}
\begin{tabular}{ccccccc}
\hline\hline
$x$ & $y$       & $m$$^{(1)}$ & $q$$^{(1)}$ & $R$$^{(1)}$ & N$_{\rm gal}$ & $X_{\rm CO}$  \\
\hline
log(M$_{\rm d}$)$_{r25}$ & log(M$_{\rm HI}$)$_{r25}$ & $0.85 \pm 0.03$ &  $3.16 \pm 0.21$ & 0.80 & 429   & --\\
log(M$_{\rm d}$)$_{r25}$ & log(M$_{\rm HI}$)$_{r25}$ & $0.85 \pm 0.04$$^{(2)}$ &  $3.07 \pm 0.30$$^{(2)}$ & 0.79$^{(2)}$ & 252$^{(2)}$ & --\\
log(M$_{\rm d}$)$_{r25}$ & log(M$_{\rm H2}$)$_{r25}$ & $1.02 \pm 0.07$ &  $1.59 \pm 0.47$ & 0.72 & 252   & const.$^{(3)}$ \\
log(M$_{\rm d}$)$_{r25}$ & log(M$_{\rm tot\,gas}$)$_{r25}$ & $0.91 \pm 0.04$ &  $3.10 \pm 0.28$ & 0.84 & 252 & const.$^{(3)}$\\
log(M$_{\rm d}$)$_{r25}$ & log(M$_{\rm H2}$)$_{r25}$ & $0.89 \pm 0.07$ &  $ 2.64\pm 0.52$ & 0.65 & 208 & $X_{\rm CO} - Z$$^{(4)}$ \\
log(M$_{\rm d}$)$_{r25}$ & log(M$_{\rm tot\,gas}$)$_{r25}$ & $0.85 \pm 0.05$ &  $3.55 \pm 0.32$ & 0.80 & 208 & $X_{\rm CO} - Z$$^{(4)}$ \\
log(M$_{\rm d}$/M$_{\rm star}$)$_{r25}$ & log(M$_{\rm HI}$/M$_{\rm star}$)$_{r25}$ & $1.21 \pm 0.05$ &  $2.74 \pm 0.15$ & 0.77 & 429       & --\\
log(M$_{\rm d}$/M$_{\rm star}$)$_{r25}$ & log(M$_{\rm HI}$/M$_{\rm star}$)$_{r25}$ & $1.28 \pm 0.06$$^{(2)}$ &  $2.95 \pm 0.20$$^{(2)}$ & 0.78$^{(2)}$ & 252$^{(2)}$       & --\\
\hline\hline
\end{tabular}
\end{adjustbox}
\tablefoot{
$^{(1)}$ Coefficients of Eq.~(\ref{eq:logm}) and Pearson coefficients $R$ of the scaling relations presented in Sect.~\ref{sec:scal-mass}.
$^{(2)}$ Dust-H{\sc i} scaling relation for sample galaxies having both H{\sc i} and H$_{2}$ data. 
$^{(3)}$ $X_{\rm CO}$ value from \citet{bolatto13}. 
$^{(4)}$ Metallicity-dependent $X_{\rm CO}$ according to the calibration of \citet{amorin16}.
}
\end{table*}

An unexpected result emerges from Figs.~\ref{fig:logmd-logmhi} and \ref{fig:logmd-logmh2totgas}: 
within the optical disk of late-type galaxies of the Local Universe, there is a tighter correlation between dust mass and atomic gas mass than between dust mass and  molecular gas mass. 
This is opposite to what is observed at smaller scales in the ISM, where dust and molecular gas 
are strongly associated in the star formation process, while the H{\sc i} gas is not directly involved with it.
In particular, there is observational evidence of a dependence of the dust--gas relation on the  ISM gas phase 
(molecular or atomic), both in the Milky Way --studied in absorption-- and in external galaxies \citep[e.g.,][]{jenkins09,roman14,chiang19}.
For instance, \citet{roman14}  found a higher DGR in the dense ISM dominated by molecular gas in the 
Magellanic Clouds observed at 10--50~pc resolution. 
Moreover, the distributions of dust, molecular gas, and total gas are rather similar and both generally  distributed in a disk, often with an exponential decline of surface density along the radius \citep[e.g.,][]{alton98,bianchi00,bigiel08,munoz09,hunt15a,casasola17,mosenkov19}.
Instead, the H{\sc i} gas does not generally follow an exponential distribution (see Sect.~\ref{sec:hi}). 
The H{\sc i} gas often has a decline towards galaxy center 
\citep[see, e.g., Figs. 3 and A.2 in][]{casasola17},
likely due to the rapid transformation  
of  the  atomic  gas  into  molecular gas, and then into stars in  the  denser central parts of galaxies \citep[e.g.,][]{bigiel08}.
The most common trend in galaxies is therefore characterized by H{\sc i} and molecular gas distributed in a complementary way. 
 
However, the weaker correlation between dust and molecular gas mass might just be the result of the uncertainties in applying a single-recipe aperture correction, 
and of the scatter around the assumed $X_{\rm CO}$ conversion factor. For instance, \citet{corbelli12} analyzed a sample of 35 metal-rich spirals of the Virgo Cluster 
and found a stronger correlation between the dust and molecular gas mass than what we find here. The reason might be that half of their objects have direct 
molecular gas masses from maps, and half estimated from aperture corrections, while in the current work the fraction of the latter is much higher ($\sim$98$\%$).
Nevertheless, our results are consistent with those of \citet{corbelli12} and 
agree with the fact that the mass of dust correlates better with the total gas mass than with the single components. An analogous result was found by \citet{orellana17}
on a sample of 189 galaxies at $z < 0.06$ with available H{\sc i} and CO observations. 

Another important point is that the assumption of a metallicity-dependent $X_{\rm CO}$  
does not lead to stronger correlations between
dust mass and molecular and total gas masses with respect to adopting a constant $X_{\rm CO}$. 
This might be due to the fact that: 
{\em i)} $X_{\rm CO}$ does not depend only on  metallicity 
(at least in our metallicity range, with 8 $<$ 12 + $\log$(O/H) $<$ 9.5~dex), and 
{\em ii)} in our sample of galaxies, spanning a range of morphologies, 
$X_{\rm CO}$ might vary along the disk due to different conditions in both metallicity, gas density, and ionizing flux, and thus our assumption of single correction is an over-simplification.

\begin{figure}[!h]
\centering
\includegraphics[width=0.51\textwidth]{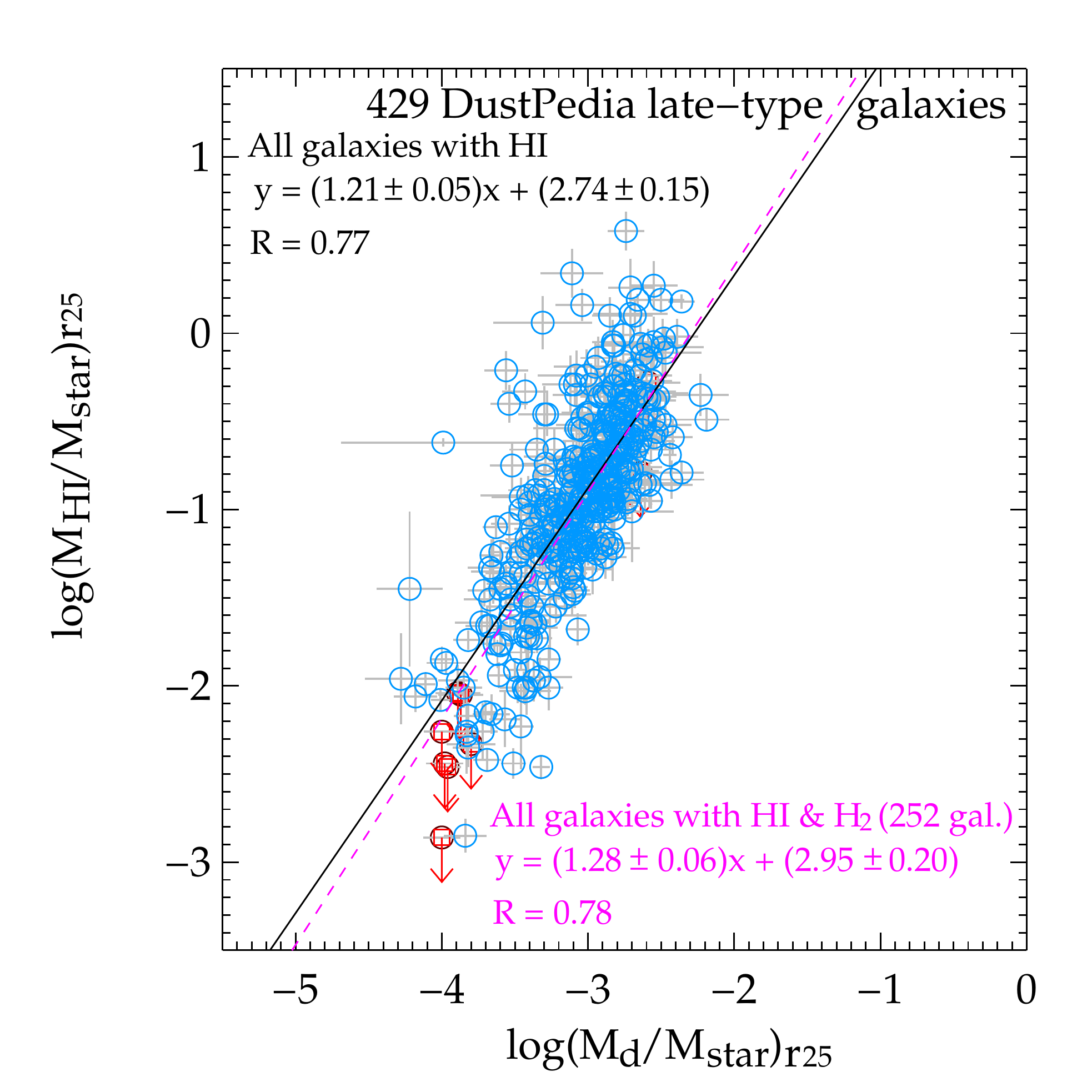}
\caption{
Same as Fig.~\ref{fig:logmd-logmhi} but with dust and H{\sc i} masses normalized with respect to the stellar mass.
}
\label{fig:dust-hi-norm-star}
\end{figure}

It is worth stressing that if we normalize the dust and gaseous masses
using the stellar mass, there is a well-defined scaling relation between the
normalized dust mass and the normalized atomic gas while there is no correlation between the normalized dust mass and the normalized molecular mass. 
The scaling relation between dust and atomic gas masses, both normalized with respect to the stellar mass, is shown in Fig.~\ref{fig:dust-hi-norm-star}.

Potentially we could compare the scaling relations between dust and gas masses with others present in the literature 
(see references above).        
However this comparison makes sense only if done coherently.
We decide to not graphically compare our ISM scaling relations with others for the following reasons:
\textit{i)} dust masses we use are the only ones obtained with the full THEMIS model that can provide dust masses up to two to three times lower
than those derived from other commonly used dust models \citep[][]{draine07};    
\textit{ii)} scaling relations involving the gas-phase metallicity are strongly dependent on the adopted calibration 
\citep[e.g., N2, $R_{23}$,][]{devis19};
\textit{iii)} we use all mass data reported to their values within $r_{25}$; and
\textit{iv)}  upper limits were treated differently, especially in gas data (e.g., survival analysis, conversion of upper limits in detections under some assumptions). 
This choice is applied to all scaling relations shown in the paper.

\begin{figure}
\centering
\includegraphics[width=0.51\textwidth]{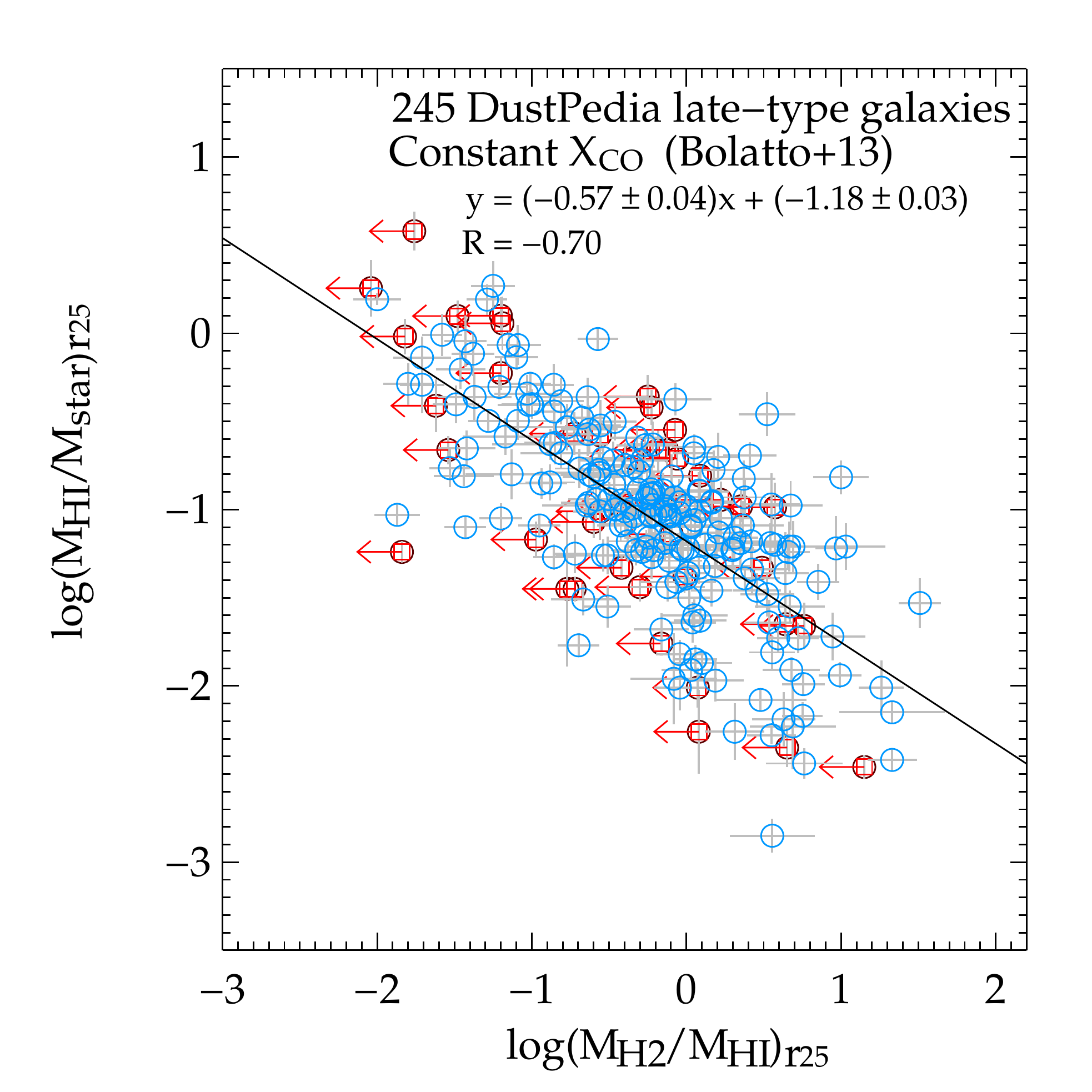}
\caption{
The H{\sc i}-to-stellar mass ratio as a function of H{\sc i}-to-H$_2$ mass ratio within $r_{25}$, in logarithmic scale, assuming a constant $X_{\rm CO}$ \citep[][]{bolatto13}. 
In this plot, we exclude data with H{\sc i} upper limits (7/252).
The black line is the fit to the drawn data.
}
\label{fig:hi_h2r25-hi_star}
\end{figure}

\subsection{Scaling relations between stellar and gas mass}
\label{sec:generalscal}
In this section, we provide a relationship between gas and stellar mass for our galaxy sample.  
The main limitation of  previous studies has been a lack of molecular gas data.
For this reason, most published studies provided gas--star relationships limited to the atomic gas component \citep[e.g.,][an exception to this is the study by Cortese~et~al.~2016]{cortese11,catinella13,gavazzi13,devis17a}.  

In Fig.~\ref{fig:hi_h2r25-hi_star}, we show H$_2$-to-H{\sc i} gas mass versus H{\sc i}-to-stellar mass ratios within $r_{25}$ under the assumption of the constant $X_{\rm CO}$.  
We find an anti-correlation between H$_2$-to-H{\sc i} and H{\sc i}-to-stellar mass ratios: 
when H$_2$ dominates over H{\sc i} (at higher H$_2$-to-H{\sc i} values), the galaxies have already formed a large 
number of stars, and thus their H{\sc i}-to-stellar mass ratio is low. 
On the other hand, galaxies dominated by H{\sc i} are less evolved, and therefore they have a lower stellar mass (a higher H{\sc i}-to-stellar mass ratio). 
The relationship  within $r_{25}$ is given by the equation:
\begin{equation}
\begin{split}
{\rm log(M_{HI}/M_{star})_{r25}}        & = (-1.18 \pm 0.03)\,+ \\ 
                                                        &+ {\rm (-0.57 \pm 0.04) \times log(M_{H2}/M_{HI})_{r25}},                                                  
\label{eq:h2-hi-star-r25}
\end{split}
\end{equation}
\noindent
assuming the constant $X_{\rm CO}$ 
($R = 0.70 $ for 245 sample galaxies).
We also derived the scaling relation of Eq.~(\ref{eq:h2-hi-star-r25}) under the assumption of a metallicity-dependent $X_{\rm CO}$, 
finding similar slope and intercept to those by assuming the constant $X_{\rm CO}$ but with lower Pearson coefficient ($R = 0.47$).    

\citet{devis19} provided a scaling relation similar to that expressed in Eq.~(\ref{eq:h2-hi-star-r25}) using the total H{\sc i} mass.
These correlations allow us to derive the H$_2$ gas mass (or the total gas mass) for a given galaxy using the H{\sc i} gas and the stellar masses.

\section{Mass ratios in the Local Universe}
\label{sec:ratios}
In this section,  we study the mass ratios between the ISM components 
and their trends as a function of galaxy morphology.
Tables~\ref{tab:sample}, \ref{tab:logfgas}, and \ref{tab:logdgr} show that all morphologies from T~=~1 to T~=~10 are well statistically represented
in various properties of the ISM.  
In our analysis we also include stellar mass. 
In particular, we study the gas fraction ($f_{\rm gas}$) defined as the ratio M$_{\rm tot\,gas}$/(M$_{\rm star}$ + M$_{\rm tot\,gas}$), 
the molecular-to-atomic gas mass ratio, the dust-to-gas mass ratios separating the various gas phases (atomic, molecular, total gas), 
and the dust-to-stellar mass ratio.
We also explore how the dust-to-total gas mass ratio varies as a function of the molecular-to-atomic gas mass ratio. 
In the following, we present and discuss these mass ratios on a case-by-case basis.
The mean values and the uncertainties of studied ratios have been derived using the survival analysis technique
based on the Kaplan-Meier estimator \citep[][]{Kaplan58}.
These quantities are collected in Tables~\ref{tab:logfgas} and \ref{tab:logdgr}.

\subsection{Gas fraction}
\label{sec:fgas}
\begin{figure}
\centering
\includegraphics[width=0.5\textwidth]{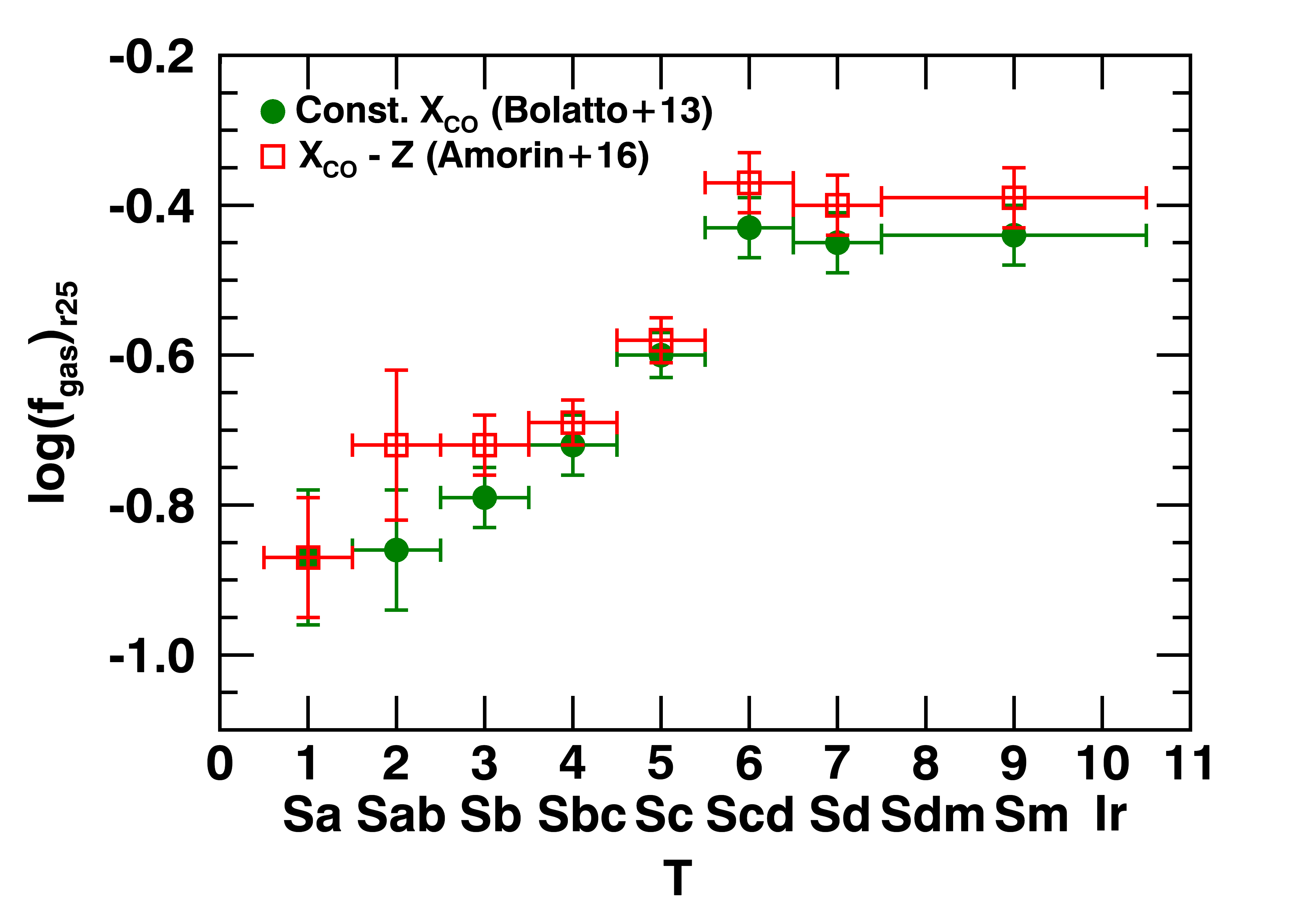}
\caption{
Gas fraction, $f_{\rm gas}$ = M$_{\rm tot\,gas}$/(M$_{\rm star}$ + M$_{\rm tot\,gas}$), within 
$r_{25}$ as a function of the morphological stage, from T~=~1 to T~=~10, assuming a constant $X_{\rm CO}$ \citep[][filled green dots]{bolatto13} and a metallicity-dependent $X_{\rm CO}$ 
\citep[][empty red squares]{amorin16}.
The symbols represent the mean values and their uncertainties (expressed in logarithm) in bins of $\Delta$T~=~1, 
except for the morphological stages T~=~8, 9, 10 that are binned together because of a smaller number of later-type galaxies
than earlier-type ones.  
}
\label{fig:logfgas}
\end{figure}

Figure~\ref{fig:logfgas} shows the trend of $f_{\rm gas}$ within $r_{25}$ as a function of the morphological stage, 
 assuming a constant and a metallicity-dependent $X_{\rm CO}$.   
The gas fraction with the metallicity-dependent $X_{\rm CO}$ is higher than that 
(but always consistent within the errors) with the constant $X_{\rm CO}$.
The gas fraction increases with increasing T from T~=~1 to T~=~6, while it remains approximately constant from T~=~6 to T~=~10.
This trend holds for both assumptions on $X_{\rm CO}$ (see Table~\ref{tab:logfgas} for mean values of $f_{\rm gas}$).

It is well known that the gas fraction is a proxy for galaxy evolution as it is a good measurement of how much star formation can be sustained from the current gas reservoir, compared to the star formation that has already happened
\citep[e.g.,][]{devis17a,devis19}.  
Inflows and outflows of gas and mergers can also affect the gas fraction \citep[e.g.,][]{casasola04,mancillas19}, and due to them,  
there is not necessarily a monotonic relation between the gas fraction and the evolutionary stage of a galaxy.
The trend shown in Fig.~\ref{fig:logfgas} is a direct consequence of the level of evolution of galaxies of different morphological type:
early galaxies (Sa-Sb) are usually more evolved than late ones (Sc-Sd-Irr) and thus they have converted a larger fraction of their baryonic mass into stars.  
On the other hand, late-type galaxies are less evolved than early-type ones since they are characterized by lower star formation efficiencies 
\citep[see, e.g.,][]{ferrini88,galli89,matteucci94,molla05,tissera05,brooks07,derossi07,mouhcine08,tassis08,magrini12}.
The approximately constant trend of $f_{\rm gas}$ as a function of galaxy morphology for T~$> 6$ galaxies might be due 
to the fact that our analysis refers to the optical disk, while later-type galaxies (e.g., the irregulars) are typically H{\sc i}-dominated and 
this gas is mainly located in the outer parts, beyond $r_{25}$ \citep[e.g. IC~10,][]{ashley14}. 

Our results on $f_{\rm gas}$ as a function of galaxy morphology are consistent with those recently 
obtained by \citet{sorai19} with the COMING\footnote{COMING is CO Multi-line Imaging of Nearby Galaxies, a 
Nobeyama Radio Observatory legacy project performed with the 45 m telescope.} survey, a project based on a sample of 344 FIR bright galaxies of all morphological types from ellipticals to irregulars.
Unlike us, \citet{sorai19} studied the ratio of the total molecular gas mass to the total baryonic mass, ignoring the atomic gas mass.
Although these latter authors found no clear trend of this fraction with the Hubble stage, it tends to decrease in early-type 
galaxies and vice versa in late-type galaxies.

\begin{figure}
\centering
\includegraphics[width=0.5\textwidth]{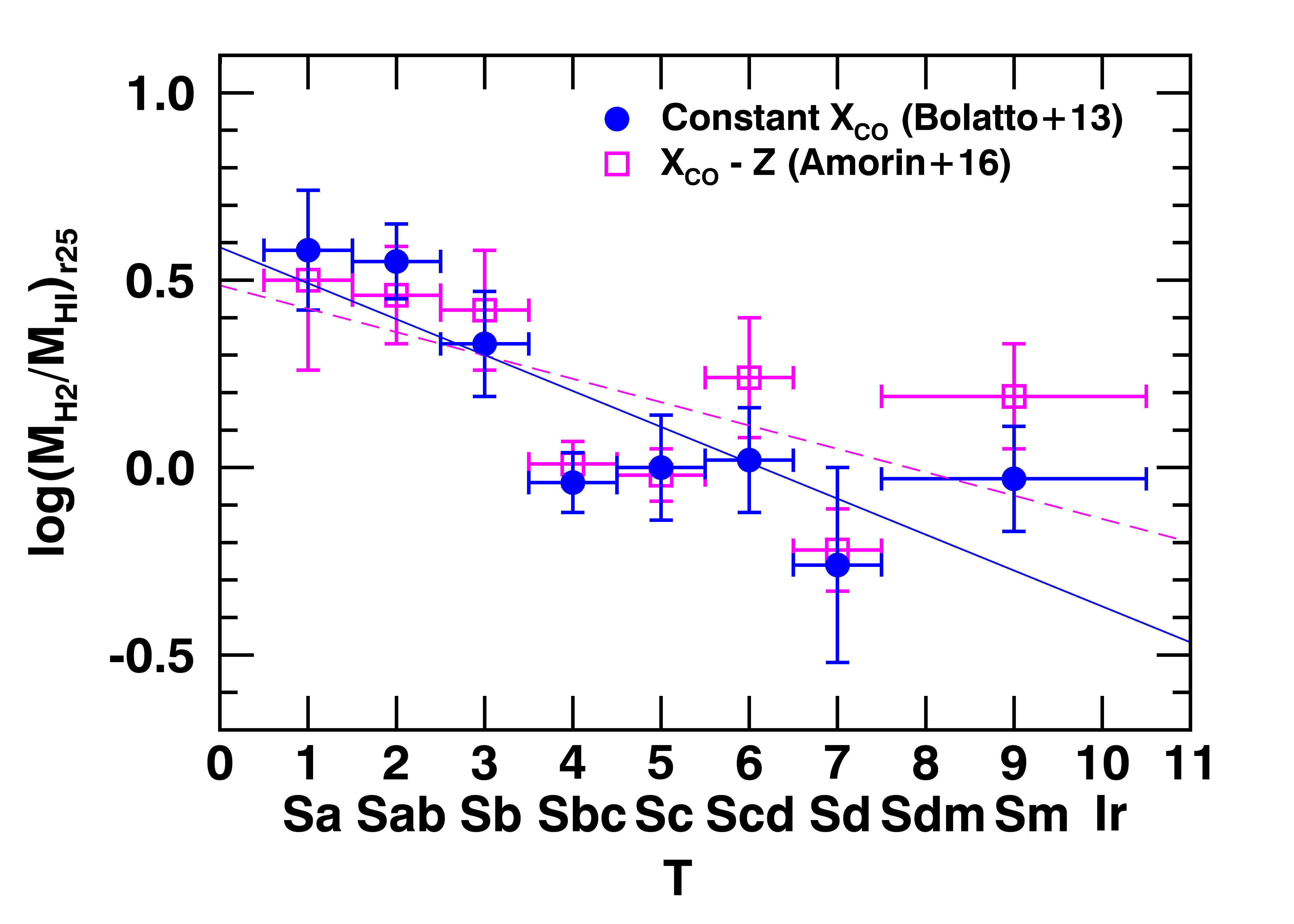}
\caption{
Molecular-to-atomic gas mass ratio within $r_{25}$ as a function of the morphological stage, from
T~=~1 to T~=~10,  assuming a constant \citep[][filled blue circles]{bolatto13} and a metallicity-dependent 
\citep[][empty magenta squares]{amorin16} $X_{\rm CO}$. 
As in Fig.~\ref{fig:logfgas}, the symbols represent the mean values and their uncertainties in bins of $\Delta$T~=~1, except for the types T~=~8, 9, 10 that are binned together. 
The (blue) continuum line is the linear fit of the data points assuming a constant $X_{\rm CO}$, the (magenta) dashed line assuming a metallicity-dependent $X_{\rm CO}$.    
}
\label{fig:logh2-hi-t}
\end{figure}

\subsection{Molecular-to-atomic gas mass ratio}
\label{sec:h2hir}
Figure~\ref{fig:logh2-hi-t} shows the variation of M$_{\rm H2}$/M$_{\rm HI}$ as a function of morphological stage within $r_{25}$,
assuming a constant and a metallicity-dependent $X_{\rm CO}$. 
Under both assumptions on $X_{\rm CO}$, M$_{\rm H2}$/M$_{\rm HI}$ tends to decrease with increasing T.
The blue continuum line and the magenta dashed line are the linear fits for a constant and a metallicity-dependent $X_{\rm CO}$
assumption, respectively, and they only help to follow the decreasing trends.
We do not provide the equation of these fits since the $x$-axis displays T that is not a numerical physical quantity 
and it depends on the adopted morphological classification of galaxies.   
The observed decreasing trend 
is again consistent with the expected galaxy evolution throughout the various morphological types:
early-type galaxies have already transformed their H{\sc i} in H$_2$, while late-type but less evolved galaxies remain dominated by atomic gas.
We refer to Table~\ref{tab:logfgas} for mean values of M$_{\rm H2}$/M$_{\rm HI}$ as a function of T.

The decreasing trend of M$_{{\rm{H}}_2}$/M$_{\rm HI}$ with increasing morphological stage is consistent with previous literature 
studies, such as that by \citet{casasola04}, both for interacting and isolated galaxies \citep[see also][]{young89,young91,bettoni03}.  
The mean values of M$_{{\rm{H}}_2}$/M$_{\rm HI}$ per T are in agreement with those of \citet{casasola04} within 2$\sigma$. 
The total mean value of M$_{{\rm{H}}_2}$/M$_{\rm HI}$ of our whole sample (log(M$_{{\rm{H}}_2}$/M$_{\rm HI}$)~$= 0.24$) is instead higher than that found by \citet{orellana17} for their sample (log(M$_{{\rm{H}}_2}$/M$_{\rm HI}$)~$=-0.10$).
This difference might be appointed to the different composition of the two samples: the dataset of  \citet{orellana17} contains a heterogeneous sample of galaxies, including different morphological types (ellipticals and spirals), isolated and interacting galaxies, 
and luminosities from normal to ultra-luminous infrared galaxies, while our database is composed of only late-type galaxies. 
In addition, \citet{orellana17} did not provide mean values of M$_{{\rm{H}}_2}$/M$_{\rm HI}$ as a function of T for comparison.

\subsection{Dust-to-gas mass ratios}
\label{sec:dgr}
\begin{figure*}
\centering
\includegraphics[width=0.495\textwidth]{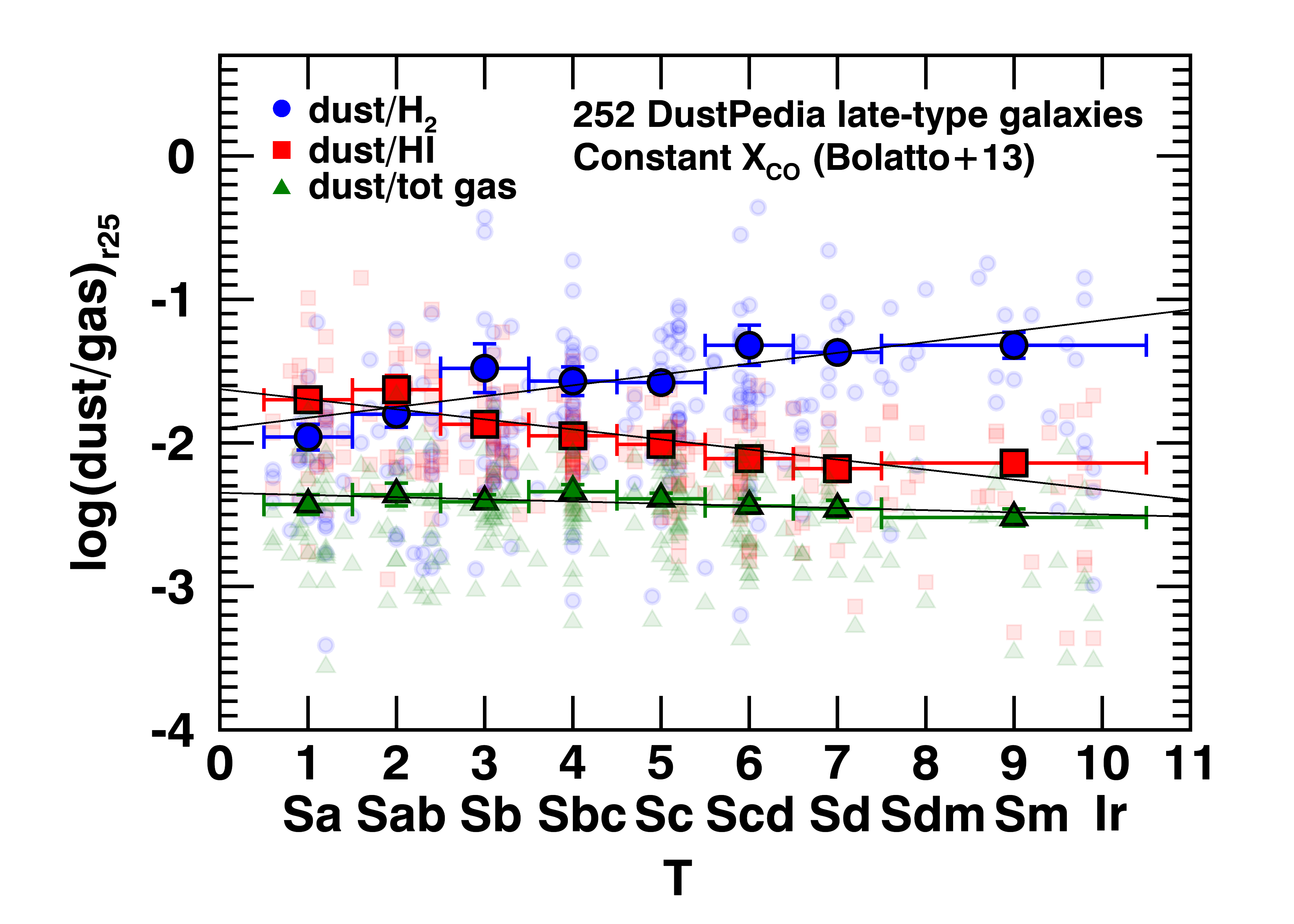}
\includegraphics[width=0.495\textwidth]{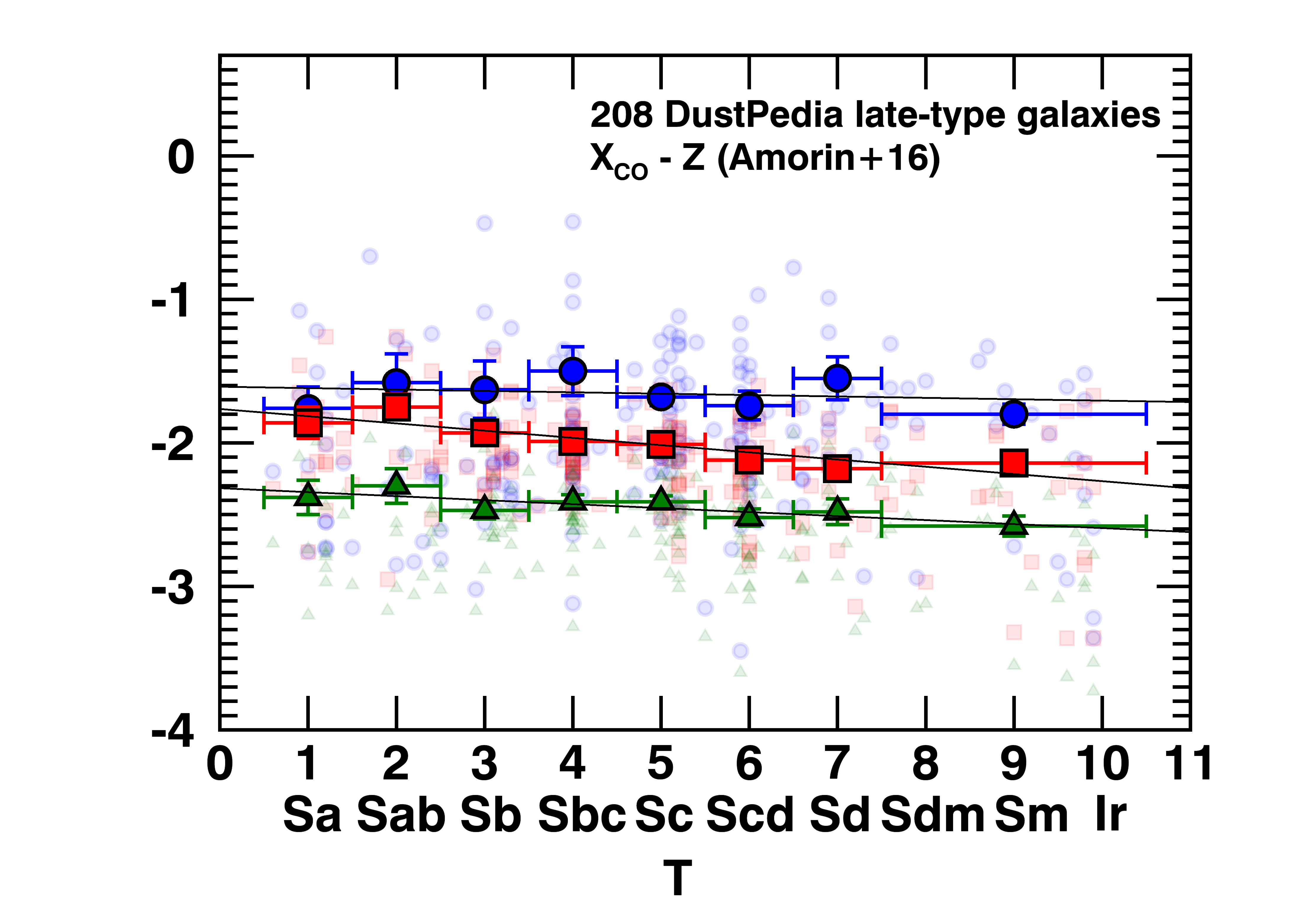}
\caption{
\textit{Left panel:} Dust-to-gas mass ratio within $r_{25}$ as a function of the morphological stage, from T~=~1 to T~=~10, assuming a constant $X_{\rm CO}$
\citep[][]{bolatto13}.
All gas types (atomic, molecular, total gas) are considered and separately plotted.
Data of individual galaxies are shown in transparency: blue circles are dust/H$_2$, red squares dust/H{\sc i}, green triangles dust/total gas.    
The large symbols represent the mean values (and their uncertainties) in bins of $\Delta$T~=~1, except for the types T~=~8, 9, 10 that are binned together.  
The black lines are the linear fits of the trends.
\textit{Right panel:} Same as \textit{Left panel} assuming a metallicity-dependent $X_{\rm CO}$ \citep{amorin16}. 
}
\label{fig:logd-g-t}
\end{figure*}

Figure~\ref{fig:logd-g-t} shows the trend of the DGR as a function of the morphological stage separating various types of gas 
(molecular, atomic, total gas), assuming a constant (\textit{Left panel}) and a metallicity-dependent (\textit{Right panel}) $X_{\rm CO}$. 
The drawn black lines are the linear fits to the data and, as in Fig.~\ref{fig:logh2-hi-t}, they only help  to follow trends and
the corresponding equations of fits are not provided.
Figure~\ref{fig:logd-g-t} shows that the dust-to-atomic gas mass ratio decreases with increasing T, and the same trend is followed by the dust-to-total gas mass ratio 
under both assumptions on $X_{\rm CO}$.
The dust-to-molecular gas mass ratio instead follows two different trends based on the assumption on $X_{\rm CO}$:
it increases with increasing T, from T~=~1 to T~=~10, when adopting a constant $X_{\rm CO}$ (\textit{Left panel}), 
while it increases from T~=~1 to T~=~4 but starts to decrease for T $>$ 4 when assuming a metallicity-dependent $X_{\rm CO}$
(\textit{Right panel}). The assumption of a metallicity-dependent $X_{\rm CO}$ better reproduces the expected decrease of the dust-to-total gas mass 
ratio \citep[e.g.,][]{draine07,remy-ruyer14,hunt15b,relano18,devis19}, as the metal content decreases with T (as is also discussed below). 
However, the trend of dust-to-H$_2$ mass ratio versus T shown in the \textit{Right panel} of Fig.~\ref{fig:logd-g-t} is mainly 
driven by the less populated bin, namely that of the late-type galaxies from Sdm to Ir.  
We refer to Table~\ref{tab:logdgr} for mean values of DGR as a function of T.

\begin{table*}
\caption{\label{tab:logfgas} 
Mean values of the gas fraction, $f_{\rm gas}$ = M$_{\rm tot\,gas}$/(M$_{\rm star}$ + M$_{\rm tot\,gas}$), and the
molecular-to-atomic gas mass ratio, within 
$r_{25}$, expressed in logarithm
as a function of the morphological stage from T~=~1 to T~=~10, and assuming 
a constant $X_{\rm CO}$ \citep[][]{bolatto13} and a
metallicity-dependent $X_{\rm CO}$ \citep[][]{amorin16}. 
Mean values of various ratios  
of all (T~=~1--10) galaxies are also collected (last line).}
\centering
\begin{adjustbox}{max width=\textwidth}
\begin{tabular}{ccccccccccccc}
\hline\hline
T & log($f_{\rm gas}$)$r_{25}$$^{(1)}$  & log($f_{\rm gas}$)$r_{25}$$^{(1)}$         & log(M$_{\rm H2}$/M$_{\rm HI}$)$r_{25}$$^{(2)}$        & log(M$_{\rm H2}$/M$_{\rm HI}$)$r_{25}$$^{(2)}$      \\
   & Const. $X_{\rm CO}$        & $X_{\rm CO}$--$Z$             & Const. $X_{\rm CO}$                                    & $X_{\rm CO}$--$Z$  \\
\hline
1 & $ -0.87 \pm 0.09 \, (26) $          & $ -0.87 \pm 0.08 \, (16)$     & $ 0.58 \pm 0.16 \, (26) $       & $ 0.50 \pm 0.24 \, (16)$ \\
2 & $ -0.86 \pm 0.08    \, (30) $               & $ -0.72 \pm 0.10 \, (16)$         & $ 0.55 \pm 0.10 \, (30) $     & $ 0.46 \pm 0.13 \, (16)$ \\
3 & $ -0.79 \pm 0.04 \, (36) $          & $ -0.72 \pm 0.04 \, (31)$     & $ 0.33 \pm 0.14 \, (36) $       & $ 0.42 \pm 0.16 \, (31)$ \\
4 & $ -0.72 \pm 0.04 \, (33) $          & $ -0.69 \pm 0.03 \, (27)$     & $ -0.04 \pm 0.08 \, (33) $      & $ 0.01 \pm 0.06  \, (27)$ \\
5 & $ -0.60 \pm 0.03 \, (41) $          & $ -0.58 \pm 0.03 \, (40)$     & $ 0.00 \pm  0.14 \, (41) $      & $ -0.02 \pm 0.07 \, (40)$  \\
6 & $ -0.43 \pm 0.04 \, (33) $          & $ -0.37 \pm 0.04 \, (31)$     & $ 0.02 \pm 0.14 \, (33) $       & $ 0.24  \pm 0.16 \, (31)$ \\
7 & $ -0.45 \pm 0.04    \, (19) $               & $ -0.40 \pm 0.04 \, (18)$         & $ -0.26 \pm 0.26 \, (19) $    & $ -0.22 \pm 0.11 \, (18)$ \\
9 & $ -0.44 \pm 0.04    \, (27) $               & $ -0.39 \pm 0.04 \, (27)$         & $ -0.03 \pm 0.14  \, (27) $   & $ 0.19  \pm 0.14 \, (27)$ \\
\hline
1--10 & $ -0.62 \pm 0.02  \,(245) $     & $ -0.55 \pm 0.02 \,(206) $    &  $0.24 \pm 0.06 \,(245) $       & $ 0.23 \pm 0.06 \,(206) $ \\
\hline\hline
\end{tabular}
\end{adjustbox}
\tablefoot{
The numbers between brackets are the numbers of galaxies taken into account in the computation of mean values.
$^{(1)}$ Values plotted in Fig.~\ref{fig:logfgas}.
$^{(2)}$ Values plotted in Fig.~\ref{fig:logh2-hi-t}. 
}
\end{table*}

\begin{figure*}
\centering
\includegraphics[width=0.495\textwidth]{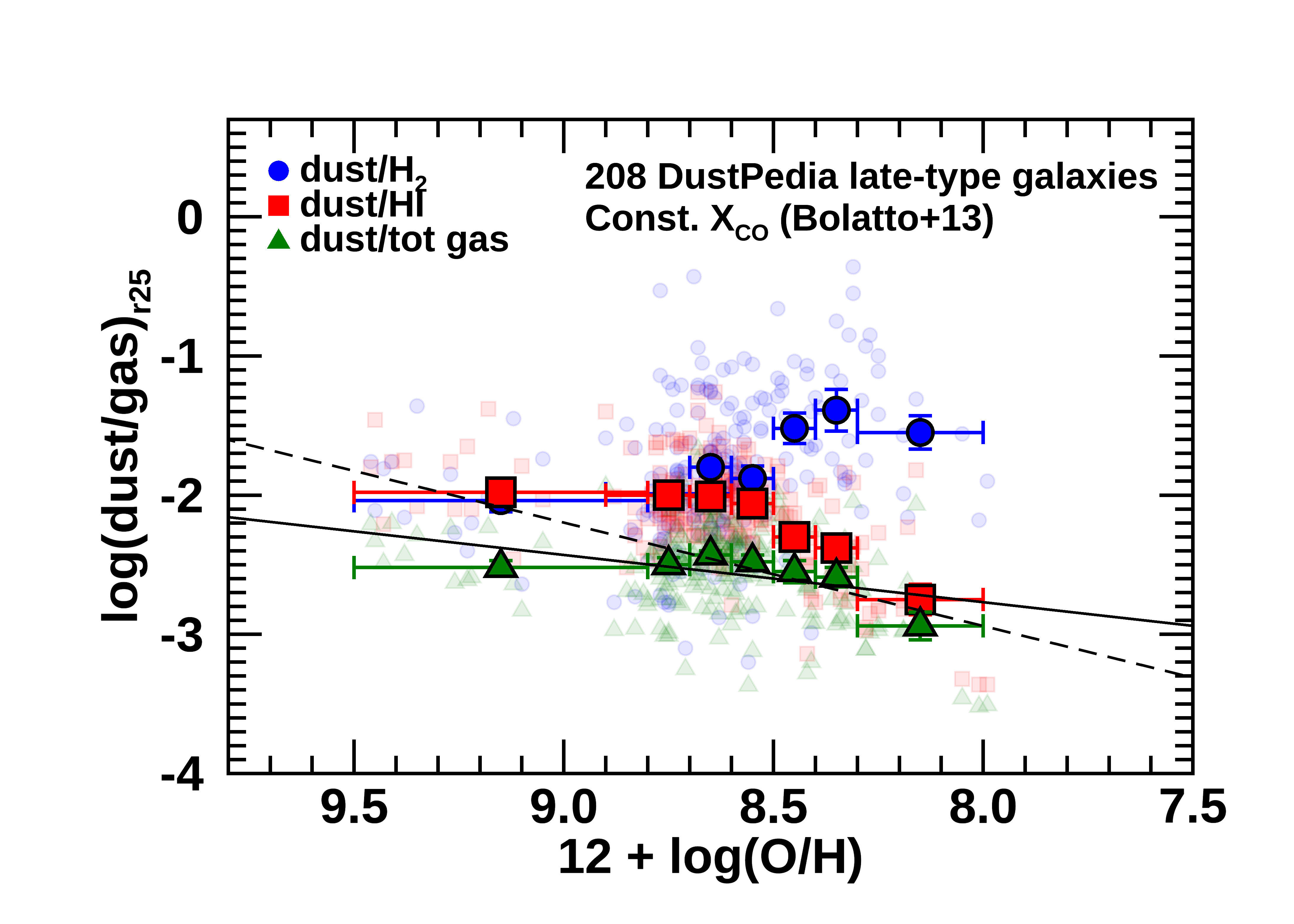}
\includegraphics[width=0.495\textwidth]{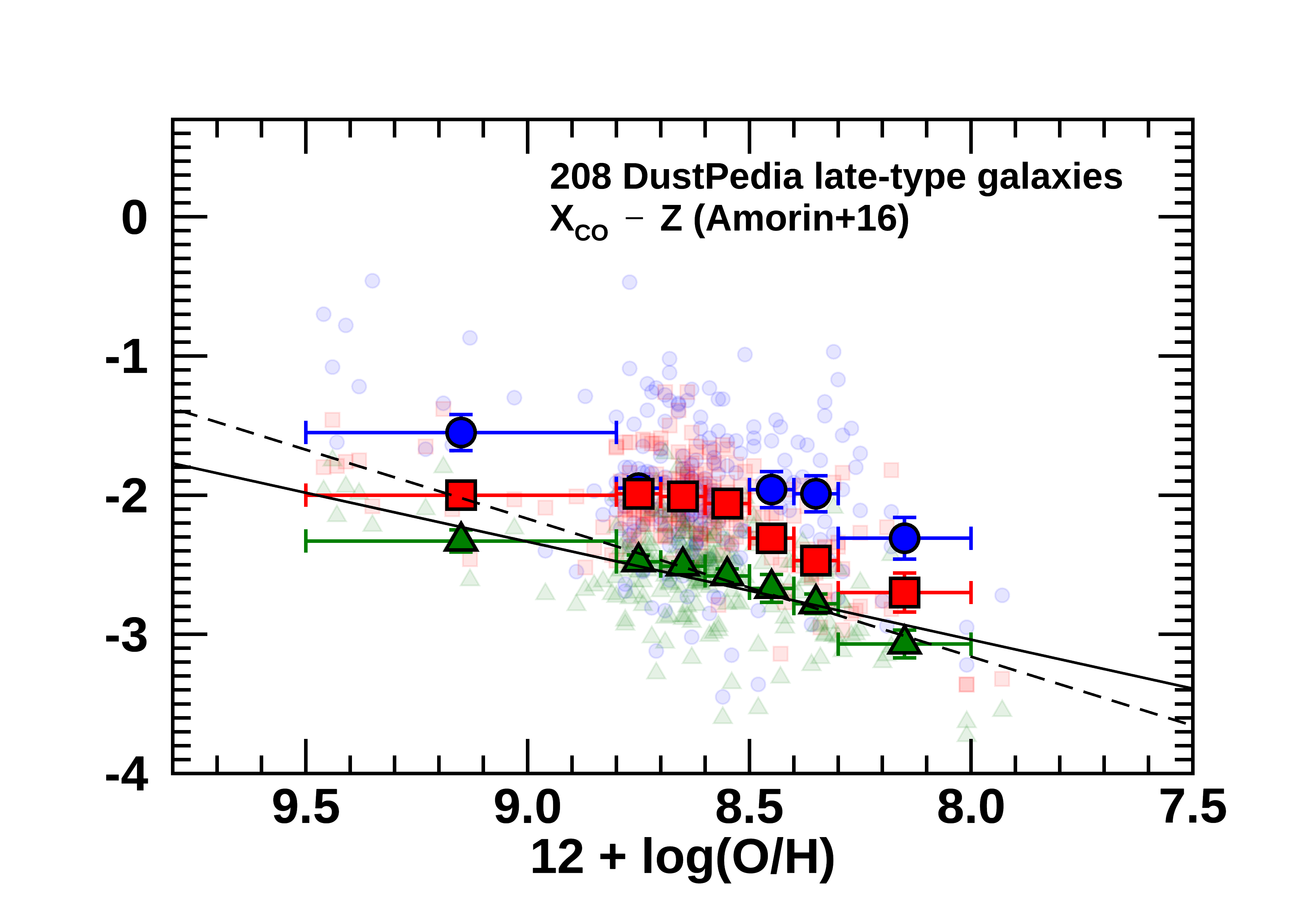}
\caption{
\textit{Left panel:} Dust-to-gas mass ratio within $r_{25}$ as a function of the gas-phase metallicity assuming a constant $X_{\rm CO}$
\citep[][]{bolatto13}.
Symbols are as in Fig.~\ref{fig:logd-g-t}.
The large symbols are mean values in metallicity bins, with the bin sizes chosen to include comparable numbers of galaxies.
The solid line is the linear fit to all the mean values of the dust-to-total gas mass ratio (big green triangles), while the dashed line is the linear fit to those excluding the high-metallicity bin.  
\textit{Right panel:} Same as \textit{Left panel} assuming a metallicity-dependent $X_{\rm CO}$ \citep{amorin16}. 
}
\label{fig:logd-g-Z}
\end{figure*}

Since the gas-phase metallicity is considered to be among the main physical properties of the galaxy driving the DGR \citep[e.g.,][]{remy-ruyer14}, 
we tested the dependence of the DGR on the gas-phase metallicity, separating the various gas phases and under both assumptions 
on $X_{\rm CO}$, as done for the morphological types.
The results of this analysis are illustrated in Fig.~\ref{fig:logd-g-Z}.
The continuum and dashed lines drawn in the two panels of this figure are the fits of the dust-to-total gas mass ratio versus metallicity taking into account
all galaxies and excluding the high-metallicity bin, respectively.
The trend of DGR as a function of metallicity is similar to 
that as a function of galaxy morphology (Fig.~\ref{fig:logd-g-t}).
In addition, the dust-to-H$_2$ mass ratio is the quantity most affected by the choice of $X_{\rm CO}$: 
while it tends to increase and decrease for low and high metallicity galaxies, respectively, with a constant $X_{\rm CO}$, 
the assumption of a metallicity-dependent $X_{\rm CO}$ realigns it with the behavior of dust-to-H{\sc i} and dust-to-total gas mass ratios. 
The coefficients of these fits are collected in Table~\ref{tab:coeff-dgr-z}.

\begin{table*}
\caption{\label{tab:coeff-dgr-z} 
Main properties of the fitting lines to DGR within $r_{25}$ as a function of 12~+~log(O/H) shown in Fig.~\ref{fig:logd-g-Z}, 
 assuming a constant and a metallicity-dependent $X_{\rm CO}$ and taking into account the total available metallicity range 
and the restricted one (neglecting the high-metallicity bin).
}
\centering
\begin{adjustbox}{max width=\textwidth}
\begin{tabular}{ccccccc}
\hline\hline
$x$ & $y$       & $m$$^{(1)}$ & $q$$^{(1)}$ & $R$$^{(1)}$ & $X_{\rm CO}$ & 12 + log(O/H)$^{(2)}$  \\
\hline
12~+~log(O/H) & log(dust/tot gas)$_{r25}$ &     $ 0.34 \pm 0.15 $ &  $-5.49 \pm 1.32 $ & 0.64 & const.$^{(3)}$ & 8.00 -- 9.50   \\
12~+~log(O/H) & log(dust/tot gas)$_{r25}$ &     $ 0.74 \pm 0.17 $ &  $-8.88 \pm 1.45 $ & 0.87 & const.$^{(3)}$ & 8.00 -- 8.90 \\
12~+~log(O/H) & log(dust/tot gas)$_{r25}$ &     $0.70 \pm 0.10 $ &  $-8.67 \pm 0.86 $ & 0.94 & $Z$-dep.$^{(4)}$ & 8.00 -- 9.50 \\
12~+~log(O/H) & log(dust/tot gas)$_{r25}$ &     $0.99 \pm 0.09 $ &  $-11.09 \pm 0.78 $ & 0.98 & $Z$-dep.$^{(4)}$ & 8.00 -- 8.90 \\
\hline\hline
\end{tabular}
\end{adjustbox}
\tablefoot{
$^{(1)}$ Slope, intercept, and Pearson correlation coefficients of the relationships shown in Fig.~\ref{fig:logd-g-Z}.  
$^{(2)}$ Considered metallicity range.
$^{(3)}$ $X_{\rm CO}$ value from \citet{bolatto13}. 
$^{(4)}$~Metallicity-dependent $X_{\rm CO}$ value according to the calibration of \citet{amorin16}. 
}
\end{table*}

From Table~\ref{tab:coeff-dgr-z} it emerges that the assumption of the constant $X_{\rm CO}$ produces a dust-to-total gas mass ratio depending 
on gas-phase metallicity as DGR~$\propto {\rm (O/H)} ^{0.3}$ (continuum line in the \textit{Left panel} of Fig.~\ref{fig:logd-g-Z}), 
not in line with theoretical expectations based on a constant dust-to-metal ratio \citep[see e.g.,][]{clark19} and predicting a slope of about $1$ (see later).
The slope becomes $0.7$ excluding the high-metallicity bin in the fit (dashed line in the \textit{Left panel} of Fig.~\ref{fig:logd-g-Z}).
On the contrary, assuming a metallicity-dependent $X_{\rm CO}$ according to the calibration of \citet{amorin16}, we find 
DGR~$\propto {\rm (O/H)} ^{0.7}$ (continuum line in the \textit{Right panel} of Fig.~\ref{fig:logd-g-Z}).
The DGR becomes $\propto {\rm (O/H)} ^{1.0}$ neglecting the high-metallicity bin in the fit  
(dashed line in the \textit{Right panel} of Fig.~\ref{fig:logd-g-Z}), which  is consistent with several theoretical expectations, mainly
those neglecting dust grain growth.
We notice that other theoretical studies including dust grain growth predict a slope steeper than 1
\citep[e.g.,][]{zhukovska14,feldmann15,aoyama17,devis17b,mcKinnon18}.    
We also mention that the slope between DGR and oxygen abundance might be slightly steeper than 1 if the full DustPedia sample is used \citep[see][and De~Vis priv. comm.]{devis19}.

Several studies have indeed shown that the DGR (also studied with the opposite ratio, the gas-to-dust mass ratio (GDR)) is well represented by a power law with a slope of 
about $1$ (and higher, or $\lesssim$$-1$ in terms of GDR) 
at high metallicities and down
to 12 + log(O/H)~$\sim$~8.0--8.2 \citep[e.g.,][]{james02,draine07,galliano08,leroy11,amorin16,giannetti17}.
\citet{remy-ruyer14}  found a GDR versus metallicity relation parametrized as
a broken power-law, with a slope of $-1$ at high-metallicity and a slope   of $-3.1 \pm 1.8$ at low-metallicity, with a transition metallicity around 8
\citep[see also][for a model accounting for the double-power-law trend between DGR and gas-phase metallicity of local galaxies]{popping17}.
Additionally, spatially resolved studies \citep[e.g.,][]{relano18,vilchez19} have found that the relation between the DGR and 12~+~log(O/H) at local scales
agrees with the general relation found for nearby galaxies by  \citet{remy-ruyer14}.
To explain the steeper trend at low metallicities, \citet{remy-ruyer14} invoked the action of the harder interstellar radiation field 
in low-metallicity dwarf galaxies that affects the balance between dust formation and destruction
by limiting the accretion or enhancing the destruction of the dust grains.
However, we stress that our sample is not characterized by very low metallicities (12~+~log(O/H)~$ = $~$7.0 - 8.0$).

Our sample extends up to 12~+~log(O/H)~=~9.5 (although only for 5.3$\%$ of the sample) and the inclusion 
of these high-metallicity galaxies has the effect of softening the slope of the DGR versus metallicity relation
over the entire metallicity range of our sample, providing a slope of $0.7$ instead of $1$ under the assumption of a metallicity-dependent $X_{\rm CO}$.
We also stress that the adopted calibration of $X_{\rm CO}$ with the metallicity from \citet{amorin16} is based on galaxy sample with 12~+~log(O/H)~$\leq$~9.0.
However, different $X_{\rm CO}$ prescriptions disagree on what happens in the high-metallicity ``uncharted territory''.
We also point out that the high-metallicity bin shown in Fig.~\ref{fig:logd-g-Z} is the largest one 
because of the smaller number of galaxies in the extreme of our sample, 
adding an uncertainty to our results.

\begin{figure}
\centering
\includegraphics[width=0.48\textwidth]{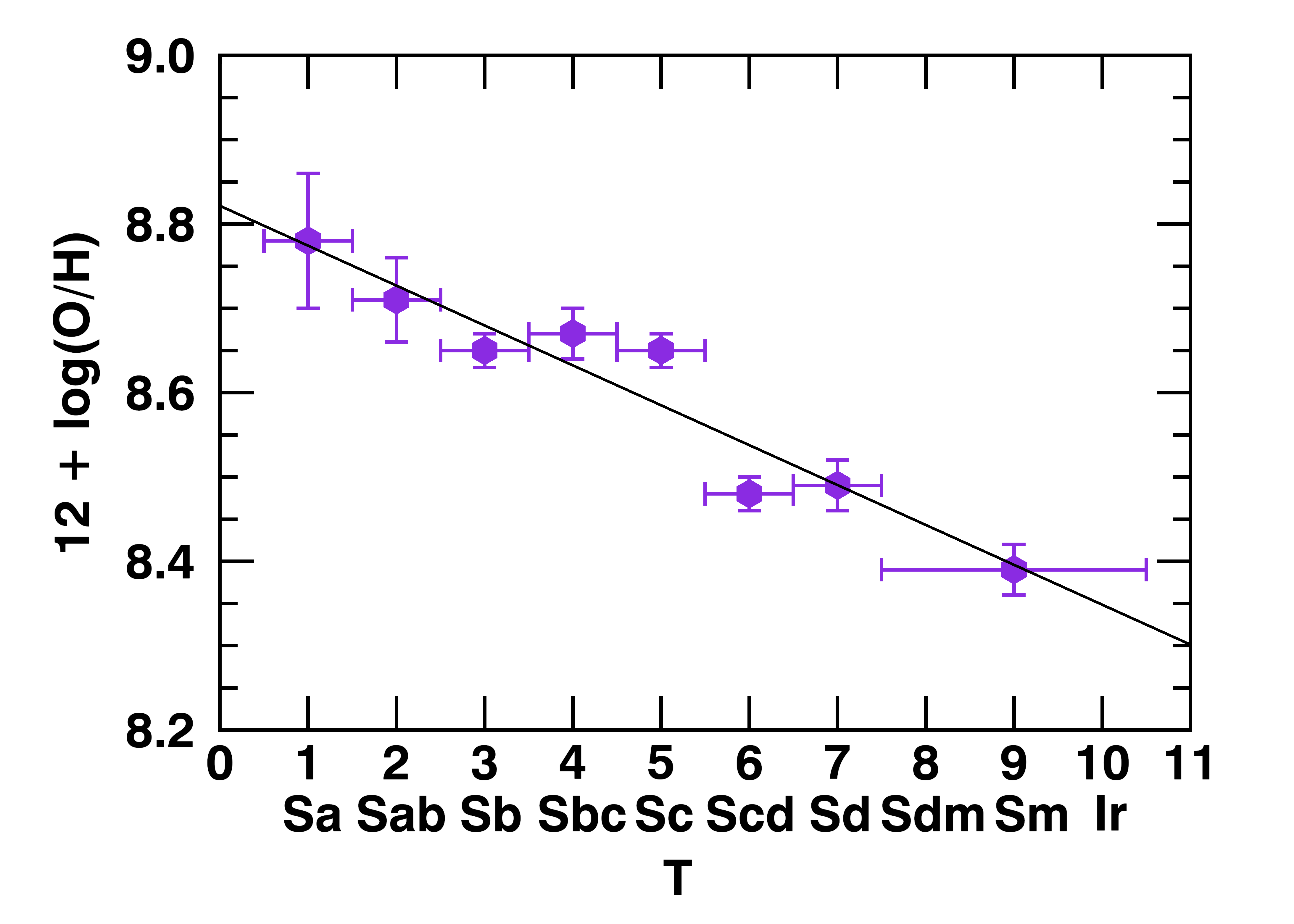}
\caption{
Oxygen abundance as a function of morphology stage, from T~=~1 to T~=~10.
As in Fig.~\ref{fig:logfgas}, 
the symbols represent the mean values (and their uncertainties) 
in bins of $\Delta$T~=~1, except for types T~=~8, 9, 10 that are binned together.  
The black line is the linear fit to the mean values.
}
\label{fig:logOH-T}
\end{figure}

To complete the characterization of our galaxy sample in terms of gas-phase metallicity, we also studied the variation 
of 12~+~log(O/H) as a function of the morphological stage. 
This analysis is complementary to that performed in Figs.~\ref{fig:logd-g-t} and \ref{fig:logd-g-Z} and allows us to better 
outline the connection between DGR, gas-phase metallicity, and galaxy morphology. 
The correlation between 12~+~log(O/H) and T is shown in Fig.~\ref{fig:logOH-T} for $\sim$78$\%$ of our sample (339 galaxies).
As expected, the oxygen abundance tends to decrease with increasing T \citep[e.g.,][]{cervantes09}.
In Table~\ref{tab:logOH-T} we have collected the mean values of 12~+~log(O/H) as a function of T for all sample galaxies 
with metallicity measurements (second column of the table) and for those with data for dust, H$_2$, and H{\sc i}
(third column of the table, i.e., for the 208 galaxies plotted in the \textit{Right panel} of Fig.~\ref{fig:logd-g-t} and in both panels of Fig.~\ref{fig:logd-g-Z}). 
We stress that values of 12~+~log(O/H) reported in the second and third columns are consistent, indicating that 
the mean metallicities for the subsample of galaxies
with complete gas and dust information are representative of the mean oxygen abundances, free of bias, 
for the whole sample.

Combining Figs.~\ref{fig:logd-g-t}, \ref{fig:logd-g-Z}, and \ref{fig:logOH-T} it emerges that DGR, 
galaxy morphology, and oxygen abundance are 
closely related properties.
By assuming the metallicity-dependent $X_{\rm CO}$, the DGR decreases with increasing morphological stage 
and with decreasing oxygen abundance. 
These two decrements are quite similar and are able to reproduce the expected trends of DGR versus T, and DGR versus 12~+~log(O/H).
We also show that the effect
of a variable $X_{\rm CO}$ mainly affects the extremes of our sample, in particular the most metal-rich bin populated by Sa-Sb galaxies, 
thus altering the global slopes of the relationships of DGR versus metallicity and DGR versus morphological type.
The most populated regions of our sample, dominated by late-type galaxies with solar metallicity, are less affected by the choice of  $X_{\rm CO}$.
Based on these findings, we suggest using a metallicity-dependent $X_{\rm CO}$, especially for studies dedicated to the 
DGR of nearby late-type galaxies.

\begin{table}
\caption{\label{tab:logOH-T} 
Mean values of the metallicity abundances derived from the N2 method \citep[][]{pettini04}, 12~+~log(O/H)$_{\rm N2}$, 
as a function of morphological stage, from T~=~1 to T~=~10.
Mean values of the metallicity abundances of all (T~=~1--10) galaxies are also collected (last line).}
\centering
\begin{adjustbox}{max width=\textwidth}
\begin{tabular}{ccc}
\hline\hline
T       & 12~+~log(O/H)$_{\rm N2}$$^{(1)}$      & 12~+~log(O/H)$_{\rm N2}$$^{(2)}$        \\
        &                                                               & with dust, H$_2$, H{\sc i} data \\
\hline
1 & $ 8.78       \pm 0.08 \, (24) $     & $ 8.79 \pm 0.09 \, (16)$ \\
2 & $ 8.71       \pm 0.05 \, (21) $     & $ 8.74 \pm 0.06 \, (16)$ \\
3 & $ 8.65       \pm 0.02 \, (47) $     & $ 8.68 \pm 0.02 \, (31)$ \\
4 & $ 8.65 \pm 0.02 \, (40) $   & $ 8.70 \pm 0.04 \, (29)$ \\
5 & $ 8.65 \pm 0.02 \, (54) $   & $ 8.65 \pm 0.02 \, (40)$ \\
6 & $ 8.48       \pm 0.02 \, (57) $     & $ 8.50 \pm 0.03 \, (31)$ \\
7 & $ 8.49 \pm 0.03 \, (34) $   & $ 8.51 \pm 0.06 \, (18)$ \\
9 & $ 8.39 \pm 0.03 \, (62) $   & $ 8.43 \pm 0.05 \, (27)$ \\
\hline
1--10 & $ 8.57 \pm 0.01    \, (339) $   & $ 8.62 \pm 0.02 \, (208)$ \\
\hline\hline
\end{tabular}
\end{adjustbox}
\tablefoot{
The numbers in brackets are the numbers of galaxies taken into account in the computation of mean values.
$^{(1)}$ Values plotted in Fig.~\ref{fig:logOH-T}.
$^{(2)}$ Values not plotted in any figure. 
}
\end{table}

\subsection{Dust-to-stellar mass ratio}
\label{sec:dsr}
\begin{figure}
\centering
\includegraphics[width=0.50\textwidth]{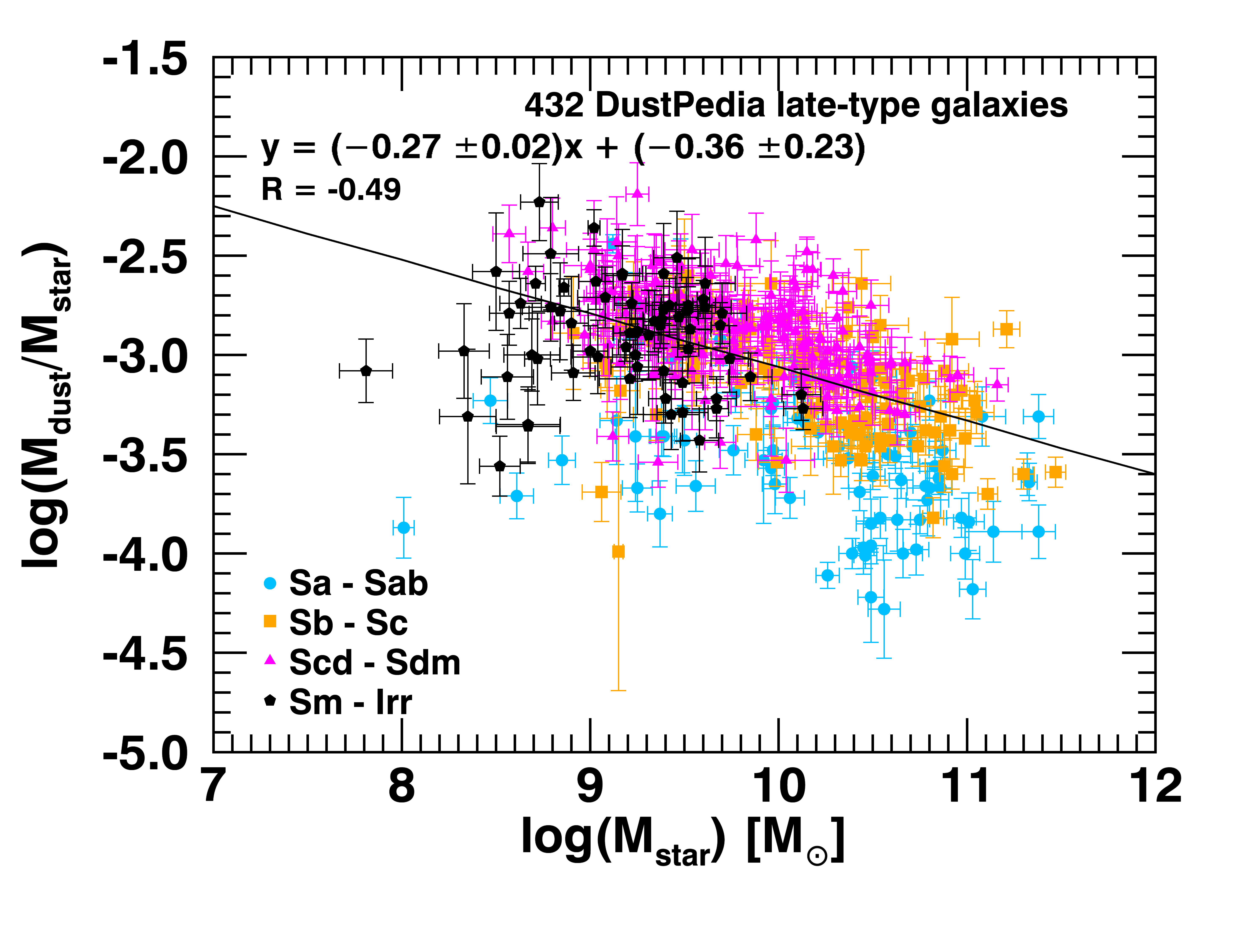}
\caption{
Dust-to-stellar mass ratio as a function of stellar mass.
Galaxies are drawn with different symbols as a function of morphological stage (four bins in galaxy morphology). 
The continuum line is the linear fit to the DustPedia late-type galaxy sample (Eq.~\ref{eq:donstar-star}).
}
\label{fig:logd_star-logstar}
\end{figure}

In Fig.~\ref{fig:logd_star-logstar}, we present the relationship between  M$_{\rm dust}$/M$_{\rm star}$ and M$_{\rm star}$,
dividing sample galaxies as a function of four main morphological stages.
There is a clear anti-correlation between M$_{\rm dust}$/M$_{\rm star}$ and M$_{\rm star}$, which is already known in the literature \citep{cortese12,clemens13,clark15,calura17,devis17a,orellana17}
and has been reproduced  by both chemical evolutionary models (see, e.g., \citealt{cortese12,calura17,devis17a}) and 
cosmological hydro simulations \citep[][]{camps16}.
The fit to our sample is given in the following equation:
\begin{equation}
\begin{split}
{\rm log(M_{dust}/M_{star})}    & = (-0.36 \pm 0.23)\,+\\
                                                &+ {\rm (-0.27 \pm 0.02) \times log(M_{star}[M_\odot])}.                         
\label{eq:donstar-star}
\end{split}
\end{equation}
\noindent

The anti-correlation displayed in Fig.~\ref{fig:logd_star-logstar} and quantified in Eq.~(\ref{eq:donstar-star}) 
suggests that galaxies with a lower stellar mass have higher dust-to-stellar mass ratios with respect to stellar mass than their more massive counterparts. 
This can be interpreted in terms of the dust lifecycle: both stars and dust are products of star formation but while stellar mass grows with time as the galaxy evolves, 
dust grains (and the metals from which they form, along with the other ISM components) eventually disappear from
the budget as they are incorporated into the stellar mass.
This also agrees with the trend of the DGR versus gas-phase metallicity displayed in Fig.~\ref{fig:logd-g-Z}.
Interestingly, \citet{calura17} found that the anti-correlation between M$_{\rm dust}$/M$_{\rm star}$ and M$_{\rm star}$ is not only valid 
for the Local Universe but continues to exist up to redshfit $z \sim 2$. 

Figure~\ref{fig:logd_star-logstar} also shows the trend in terms of morphological distribution.
As already noticed by \citet{cortese12}, the dust-to-stellar mass ratio monotonically decreases when moving from late-
to early-type galaxies.
\citet{dacunha10} explain how this trend is in line with the decreasing  specific SFR (sSFR, SFR per unit of M$_{\rm star}$) 
with increasing M$_{\rm star}$ \citep[e.g.,][]{schiminovich07,catinella18}. 
Although the study of the sSFR is beyond the scope of this work, it is worth reiterating that low-stellar-mass late-type galaxies 
have typically high sSFR and gas fraction.
The gas is able to fuel the star formation, and subsequently large amounts of dust are formed.   
For high-stellar-mass earlier-type galaxies, the sSFR and gas fraction instead tend to decrease, and consequently the production of dust 
is lower and likely not able to compensate for or exceed that destroyed in the ISM during these processes.       
Figure~\ref{fig:logd_star-logstar} also shows that Sa-Sab galaxies are almost all below the line fit but tend to cover the entire range 
of M$_{\rm star}$. 
This could explain the diagonal scatter around the fit line.
Sb-Sdm galaxies instead appear to lie more coherently around the fit line, and Sm-Irr galaxies, although they tend to follow the general trend, contribute to the diagonal scatter around the fit line with galaxies below it, especially those with lower M$_{\rm star}$. 

\begin{figure}
\centering
\includegraphics[width=0.5\textwidth]{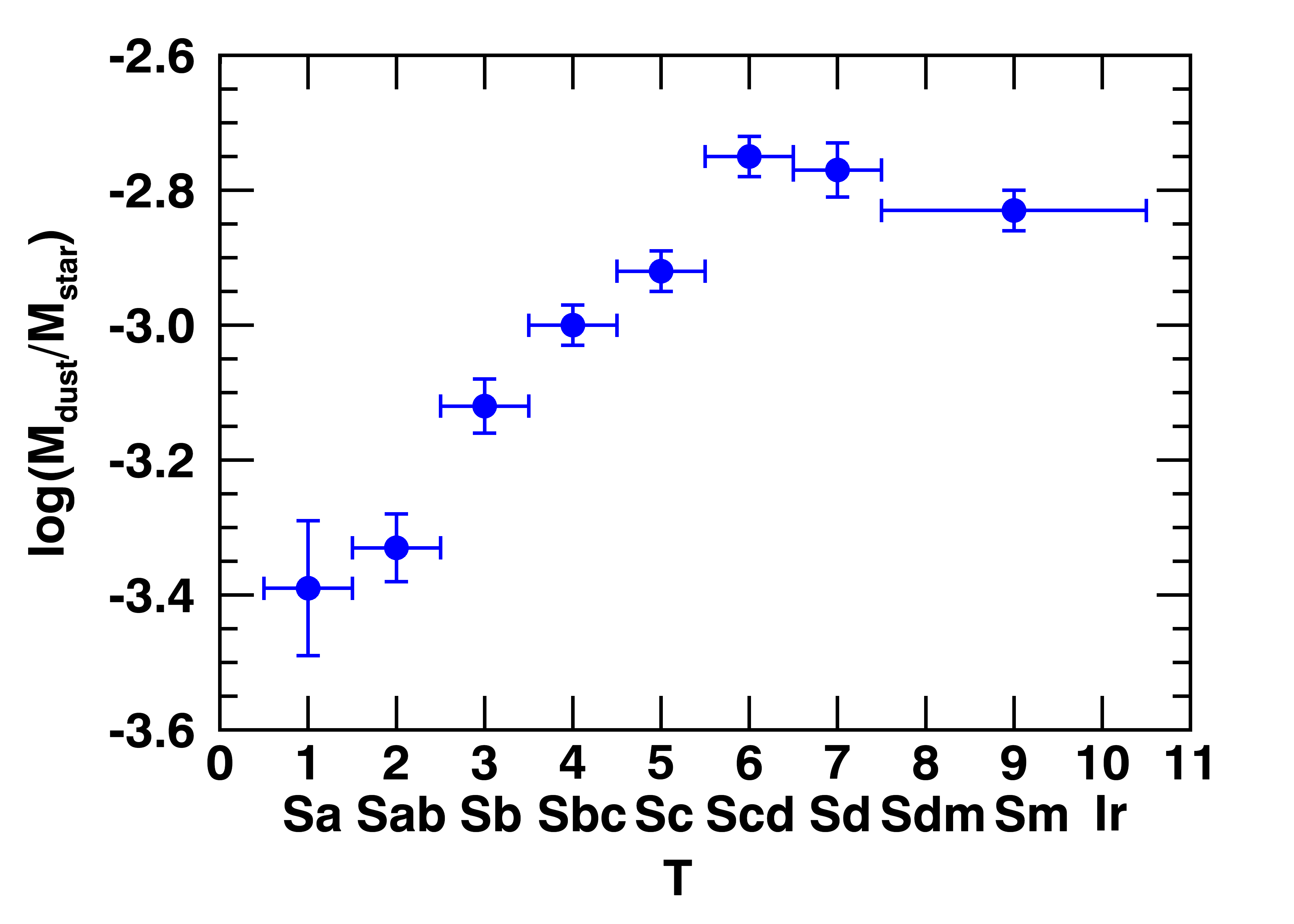}
\caption{
Dust-to-stellar mass ratio as a function of the morphological stage, from T~=~1 to T~=~10.
As in Fig.~\ref{fig:logfgas} the symbols represent the mean values (and their uncertainties) 
in bins of $\Delta$T~=~1, except for types T~=~8, 9, 10 that are binned together.  
}
\label{fig:logd-star-t}
\end{figure}

Figure~\ref{fig:logd-star-t} shows the trend of log(M$_{\rm dust}$/M$_{\rm star}$) as a function of morphological stage. 
The dust-to-stellar mass ratio increases with increasing T, from T~=~1 to T~=~6, and remains approximately constant within the
uncertainties (or slightly decreases) from T~=~6 to T~=~10. 
This trend is similar to that found for $f_{\rm gas}$ as a function of T (see Fig.~\ref{fig:logfgas}), suggesting -together 
with scaling relations- that the dust mass, total gas mass, stellar mass, and morphology of galaxies might be directly linked.
This aspect has  already been pointed out by \citet{cortese12}, although restricted to the atomic gas instead of the total gas. 
However, unlike us,  \citet{cortese12} found that M$_{\rm dust}$/M$_{\rm star}$ continues rising up to Sm (T~=~9) galaxies, 
although this increase is softer than that of earlier-type galaxies.
This disagreement could be due to the different adopted dust models and/or different treatment of data.  
Table~\ref{tab:logdgr} presents the mean values of the dust-to-stellar mass ratio as a function of morphological stage.\\

\begin{sidewaystable*}
\caption{\label{tab:logdgr} 
Mean values of the dust-to-gas mass ratio (separating molecular, atomic, and total gas) and
dust-to-stellar gas mass ratio within $r_{25}$, in logarithmic scale, 
as a function of the morphological stage, from T~=~1 to T~=~10.
For the molecular gas, we assume both a constant $X_{\rm CO}$ \citep[][]{bolatto13} and a
metallicity-dependent $X_{\rm CO}$ \citep[][]{amorin16}.
Mean values of the ratios of all (T~=~1--10) galaxies are also collected (last line).}
\centering
\begin{adjustbox}{max width=\textwidth}
\begin{tabular}{ccccccccccccc}
\hline\hline
T 
& log(M$_{\rm dust}$/M$_{\rm H2}$)$r_{25}$$^{(1)}$ & log(M$_{\rm dust}$/M$_{\rm H2}$)$r_{25}$$^{(1)}$ & log(M$_{\rm dust}$/M$_{\rm HI}$)$r_{25}$$^{(1)}$ 
& log(M$_{\rm dust}$/M$_{\rm HI}$)$r_{25}$$^{(1)}$ & log(M$_{\rm dust}$/M$_{\rm HI}$)$r_{25}$$^{(2)}$ & log(M$_{\rm dust}$/M$_{\rm tot\,gas}$)$r_{25}$$^{(1)}$
& log(M$_{\rm dust}$/M$_{\rm tot\,gas}$)$r_{25}$$^{(1)}$         & log(M$_{\rm dust}$/M$_{\rm star}$)$^{(3)}$  \\
& Const. $X_{\rm CO}$   & $X_{\rm CO}$--$Z$             
& with H$_2$ & with H$_2$ and $Z$ & & Const. $X_{\rm CO}$       & $X_{\rm CO}$--$Z$ \\
\hline
1 
& $-1.96 \pm 0.09 \, (31) $      
& $-1.76 \pm 0.15 \, (16)$  
& $-1.70 \pm 0.08 \, (31) $ 
& $-1.86 \pm 0.11 \, (16)$ 
& $-1.75 \pm 0.07 \, (44)$ 
& $-2.43 \pm 0.07 \, (31)$
& $-2.38 \pm 0.12 \, (16)$ 
& $-3.39 \pm 0.10 \, (45)$
\\
2 
& $-1.80 \pm 0.09  \,  (30) $
& $-1.58 \pm 0.20  \, (16)$ 
& $-1.63 \pm 0.10 \,  (30) $ 
& $-1.75 \pm 0.09 \, (16)$ 
& $-1.62 \pm 0.08 \, (36)$ 
& $-2.36 \pm 0.08 \, (30)$
& $-2.30 \pm 0.12 \, (16)$ 
& $-3.33 \pm 0.05 \, (37)$ 
\\
3 
& $-1.48 \pm 0.09 \, (36) $
& $-1.63 \pm 0.20 \, (31)$ 
& $-1.87 \pm 0.06 \, (36) $ 
& $-1.93 \pm 0.06 \, (31)$
& $-1.95 \pm 0.05 \, (59)$ 
& $-2.41 \pm 0.05 \, (36)$
& $-2.47 \pm 0.06 \, (31)$ 
& $-3.12 \pm 0.04  \, (59)$
\\
4 
& $-1.57 \pm 0.17 \, (35) $
& $-1.50 \pm 0.17 \, (29)$ 
& $-1.95 \pm 0.04 \, (35) $       
& $-1.99 \pm 0.05 \, (29)$ 
& $-2.00 \pm 0.04 \, (56)$ 
& $-2.34 \pm 0.05 \, (35)$
& $-2.41 \pm 0.05 \, (29)$ 
& $ -3.00 \pm 0.03 \, (56)$
\\
5 
& $-1.58 \pm 0.06 \, (41) $
& $-1.68 \pm 0.16 \, (40)$ 
& $-2.01 \pm 0.04 \, (41) $ 
& $-2.01 \pm 0.05 \, (40)$  
& $-2.03 \pm 0.03 \, (61)$ 
& $-2.39 \pm 0.04 \, (41)$
& $-2.41 \pm 0.04 \, (40)$ 
& $-2.92 \pm 0.03  \, (61)$
\\
6  
& $-1.32 \pm 0.14 \, (33) $
& $-1.74 \pm 0.10 \, (31)$ 
& $-2.11 \pm 0.05 \, (33) $       
& $-2.12 \pm 0.05 \, (31)$ 
& $-2.08 \pm 0.04 \, (70)$ 
& $-2.44 \pm 0.05 \, (33)$
& $-2.52 \pm 0.06  \, (31)$ 
& $-2.75 \pm 0.03  \, (70)$
\\
7 
& $-1.37 \pm 0.02 \, (19) $
& $-1.55 \pm 0.15 \, (18)$ 
& $-2.18 \pm 0.09 \, (19) $       
& $-2.18 \pm 0.07 \, (18)$ 
& $-2.15 \pm 0.05 \, (37)$ 
& $-2.46 \pm 0.06 \, (19)$
& $-2.48 \pm 0.09 \, (18)$ 
& $-2.77 \pm 0.04  \, (37)$
\\
9 
& $-1.32 \pm 0.09 \, (27) $
& $-1.80 \pm 0.07  \, (27)$ 
& $-2.14 \pm 0.07 \, (27) $ 
& $-2.14 \pm 0.07 \, (27)$ 
& $-2.17 \pm 0.04 \, (66)$ 
& $-2.52 \pm 0.06 \, (27)$
& $-2.58 \pm 0.07 \, (27)$ 
& $-2.83 \pm 0.03 \, (67)$
\\
\hline
1--10 
& $-1.52  \pm 0.09   \,(252) $  
& $-1.65 \pm 0.06 \,(208) $ 
& $-1.90  \pm 0.03   \,(252) $ 
& $-1.99 \pm 0.02 \,(208) $ 
& $-1.96 \pm 0.02 \,(429) $
& $-2.41 \pm 0.02   \,(252) $ 
& $-2.45  \pm 0.03   \,(208) $   
& $-2.94 \pm 0.02  \, (432)$
\\
\hline\hline
\end{tabular}
\end{adjustbox}
\tablefoot{
The numbers between brackets are the numbers of galaxies taken into account in the computation of mean values.
$^{(1)}$ Values plotted in Fig.~\ref{fig:logd-g-t}.
$^{(2)}$ Values not plotted in any figure. 
$^{(3)}$ Values plotted in Fig.~\ref{fig:logd-star-t}.
}
\end{sidewaystable*}

\begin{figure*}[!ht]
\centering
\includegraphics[width=0.49\textwidth]{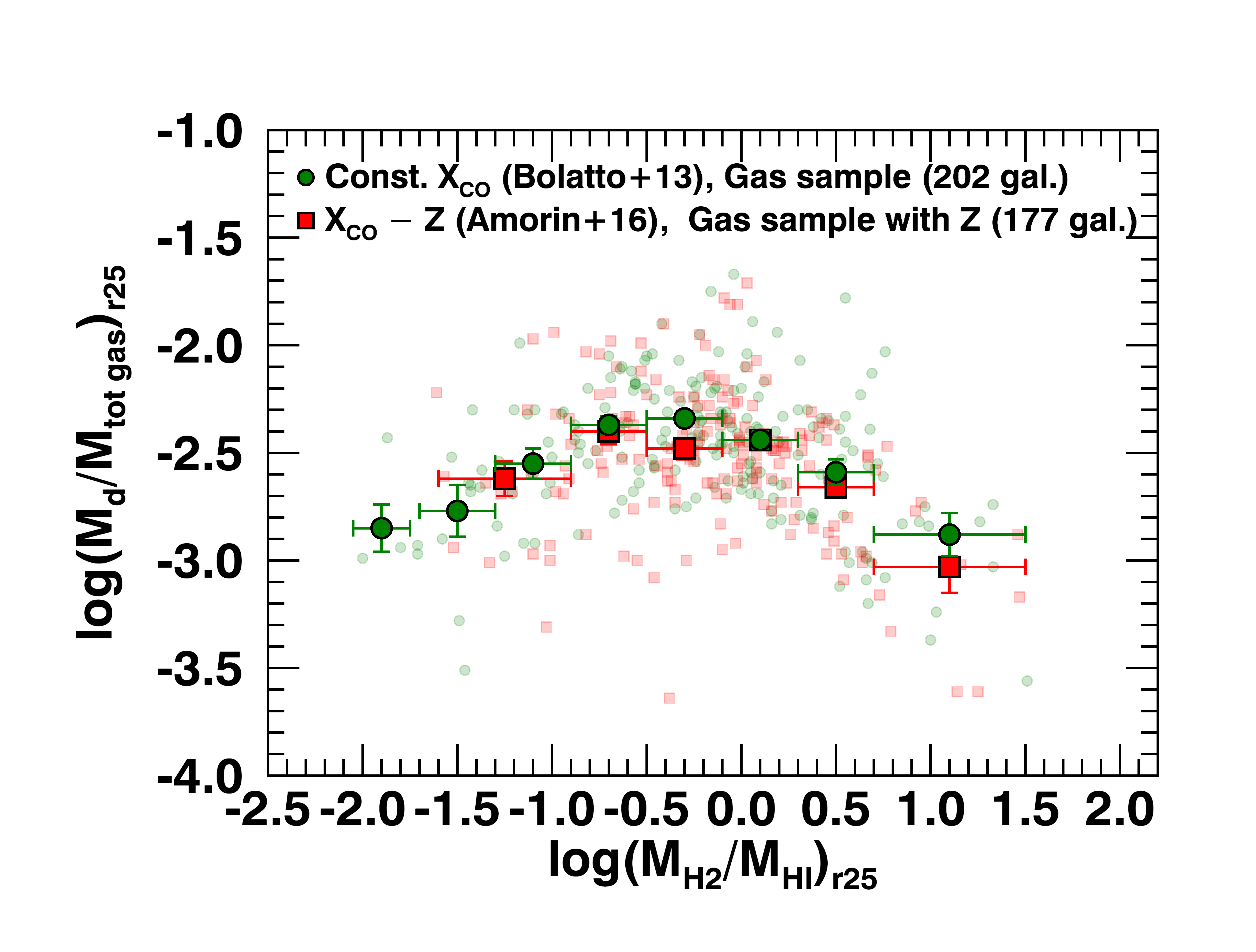}
\includegraphics[width=0.49\textwidth]{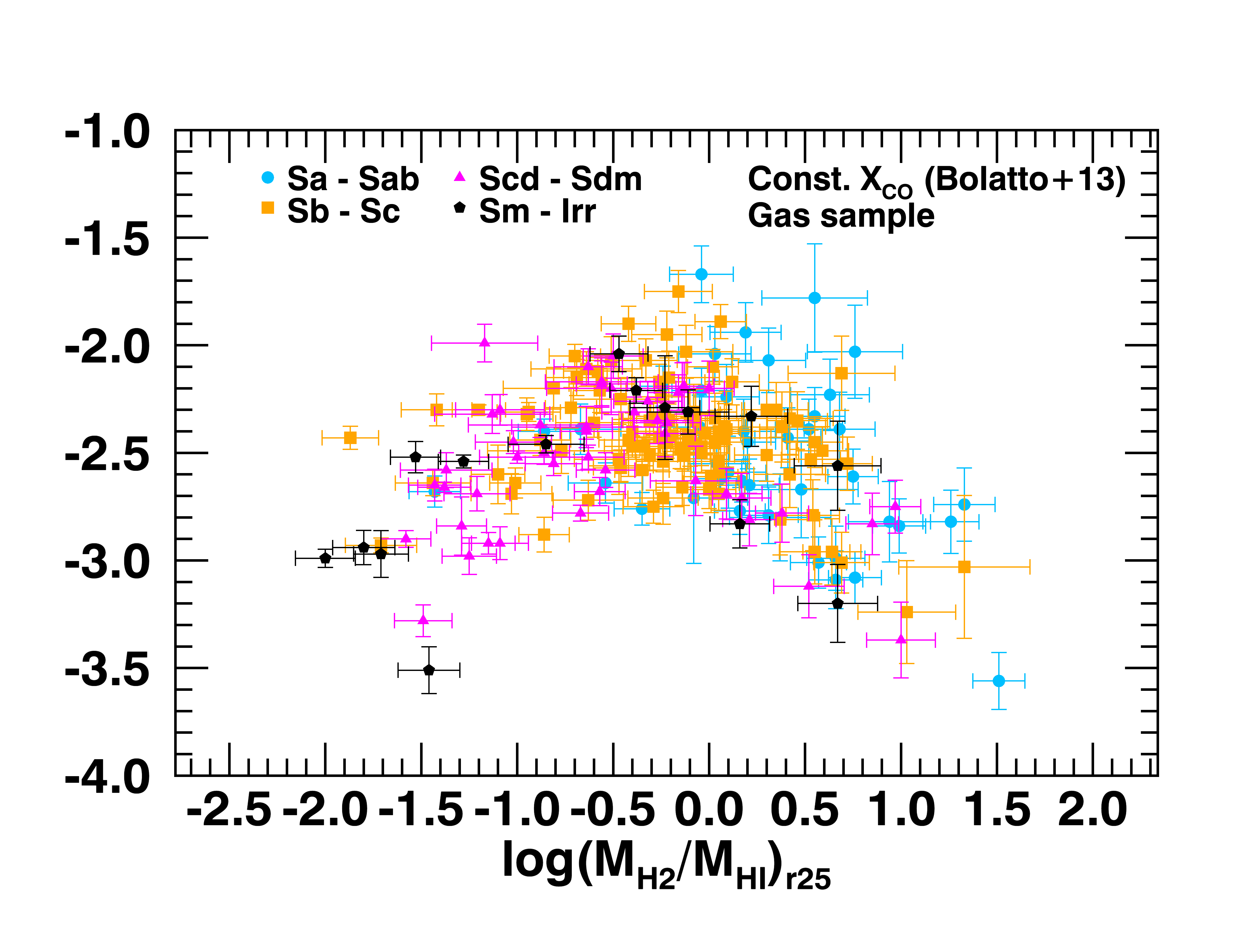}
\caption{
{\it Left panel:}
Dust-to-total gas mass ratio as a function of the molecular-to-atomic gas mass ratio within $r_{25}$ for the gas sample.
The small, transparent green circles and red squares are data points assuming a constant and a metallicity-dependent $X_{\rm CO}$, respectively.    
These data points are drawn without error bars.
The large symbols are the means of $y$-axis values in bins of $x$-axis values.
{\it Right panel:}
Same as \textit{Left panel}, under the assumption of the constant $X_{\rm CO}$ and  
displaying data points as a function of morphological stage (four main bins in galaxy morphology). 
}
\label{fig:hill}
\end{figure*}

\subsection{Dust-to-total gas mass ratio versus molecular-to-atomic gas mass ratio}
\begin{figure}
\includegraphics[width=0.5\textwidth]{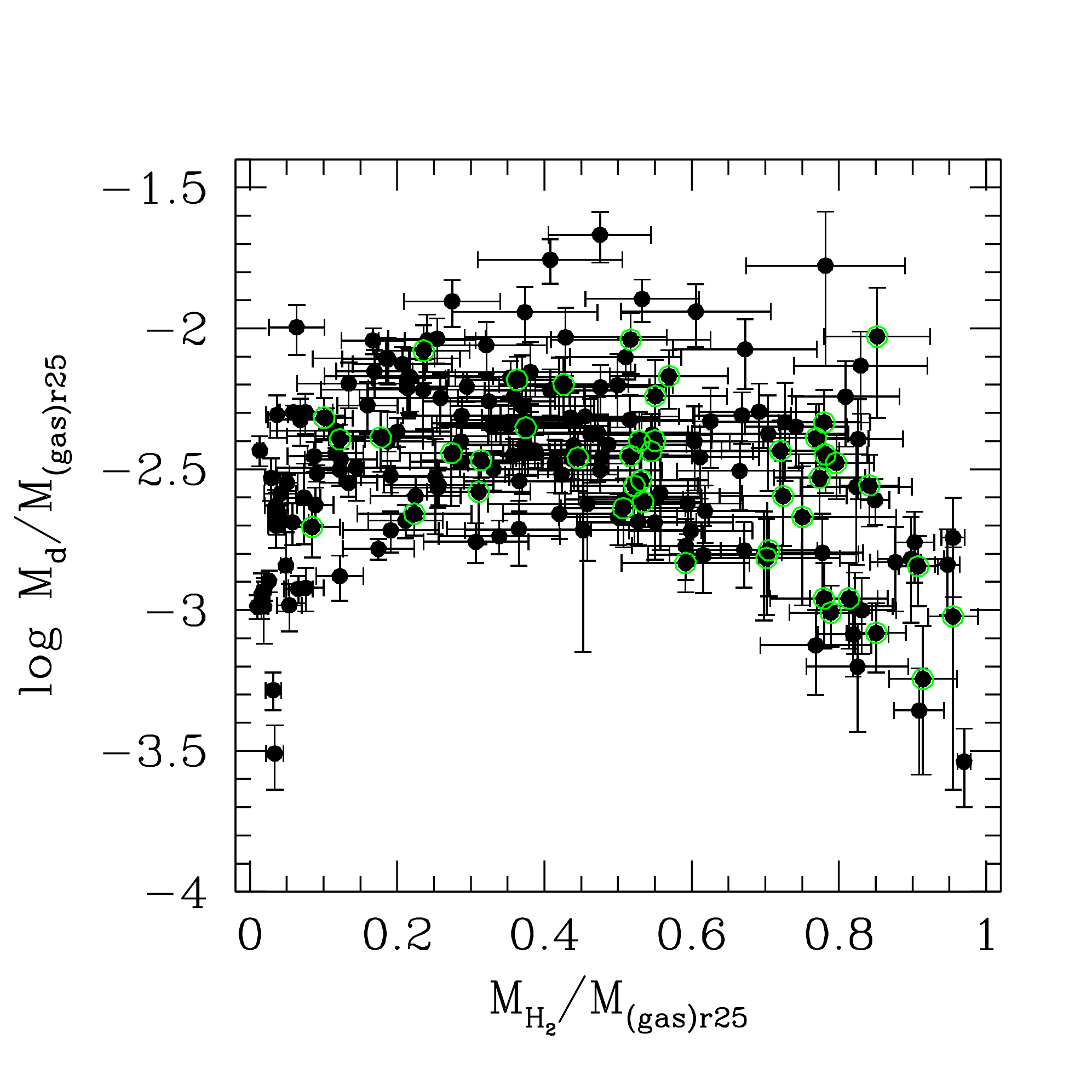}
\caption{Dust-to-total gas mass ratio as a function of the molecular fraction within $r_{25}$ is shown for the gaseous sample.  
A constant $X_{\rm CO}$ has been used. 
The open (green) circles underline active galaxies in the sample.
}
\label{fig:fh2}
\end{figure}

We also explored a new --to our knowledge-- combination of mass ratios.
Having a well-defined gaseous sample with both H{\sc i} and H$_2$ data over more than two orders of magnitude,
we can study the dust-to-total gas mass ratio as a function of the molecular-to-atomic gas mass ratio within $r_{25}$, 
or equivalently, as a function of the fraction of gas in molecular form $f_{\rm {mol}}$ (= M$_{\rm H2}$/(M$_{\rm H2}$ + M$_{\rm HI}$), or
(1.36~$\times$~M$_{\rm H2}$)/M$_{\rm tot\,gas}$ if the total gas mass includes the helium contribution, as in the present work). 
We want to stress that though the relation DGR versus H$_2$/H{\sc i} mass ratio might not have been published so far, the dependence of the DGR on the ISM gas phase (or gas density) has  previously been explored as already mentioned in Sect.~\ref{sec:scal-mass} \citep[e.g.,][]{jenkins09,roman14,chiang19}.
The results of this analysis are shown in Figs.~\ref{fig:hill} and \ref{fig:fh2} and refer to the gaseous sample.
From these figures it emerges that data are distributed along a hill-like shape.

First, we note that the shape of the distribution is mostly independent of how we model  
the $X_{\rm CO}$ conversion factor (\textit{Left panel} of Fig.~\ref{fig:hill}). 
Therefore, the downward side for high M$_{\rm H2}$/M$_{\rm HI}$ is not biased by the larger uncertainties on 
$X_{\rm CO}$ for the high-metallicity earlier-type galaxies that populate this part of the plot (see Fig.~\ref{fig:logOH-T} and later in this section).
The upward-sloping side of the hill-like distribution cannot be traced properly using a metallicity-dependent $X_{\rm CO}$ because 
of the lack of metal-abundance determinations in galaxies with a low M$_{\rm H2}$/M$_{\rm HI}$. 
Therefore, in the following discussion, for the sake of simplicity, we discuss the results obtained with a constant $X_{\rm CO}$.

Assuming a constant $X_{\rm CO}$ for the whole galaxy sample, we would expect a flat slope of the DGR versus M$_{\rm H2}$/M$_{\rm HI}$, 
within the errors.
On the other hand, we expect both M$_{\rm H2}$/M$_{\rm HI}$ and DRG to increase in more evolved galaxies since atomic gas is transformed 
into molecular gas and then into stars. 
The presence of stars  increases the hydrostatic pressures in the disk, favoring the formation of 
denser molecular clouds \citep[e.g.,][]{elmegreen93,krumholz09,wong13}. 
In addition, during the latest phases of their evolution, stars pollute the ISM with metals and dust. 
Thus, while the increasing trend on the left-side of the plot is in line with the above-mentioned expectations, the right-most part of the plot
is instead more puzzling, showing a decrease of DGR for the more evolved/molecular-gas-dominated galaxies.

Since the gas masses are used in the quantities plotted on both axes, we checked the influence of the errors 
(and in particular those on molecular gas mass) in the trends we detected. 
With a Monte Carlo procedure, we derived mock values for the gas and dust masses: for atomic and molecular gas, 
the random values are obtained from the true values and their uncertainty; for dust, we assumed a constant DGR 
(we used the average value at the peak of the hill-like distribution curve) and derived random values from the true total gas mass and 
the average uncertainty on M$_{\rm dust}$. 
The average trend for this ``null hypothesis'', obtained after 1000 representations of the dataset,
provides an approximatively constant DGR as a function of M$_{\rm H2}$/M$_{\rm HI}$. 
Indeed the large errors can produce a downward trend for high M$_{\rm H2}$/M$_{\rm HI}$, but never to the level 
we detected in the true dataset, and therefore the observed trend is real.

In the \textit{Right panel} of Fig.~\ref{fig:hill}, we investigate how the different morphological types are distributed over the hill-like curve. 
While galaxies with intermediate morphological type from Sb to Sd are located all across the curve,  
Sm and Ir galaxies are predominantly located towards to left-side of the plot where H{\sc i} is the dominant gas phase and 
the dust is still in small quantities with respect to the total gas, and the early-type Sa and Sab are located on the right-side
where the H$_2$ component dominates. 
The galaxy evolution highlighted by the \textit{Right panel} of Fig.~\ref{fig:hill} is consistent with that of Fig.~\ref{fig:logd_star-logstar}, and confirms that galaxies with higher H$_2$/H{\sc i} mass ratios are bulge-dominated systems 
as found, for instance, with COLD GASS and xGASS surveys \citep[e.g,][]{saintonge16,catinella18}. 

The {unexpected} decrease of DGR in H$_2$-dominated galaxies can be due to the interplay of different effects: 

{\em i)} Dust might be destroyed by strong stellar radiation fields or by shocks at a rate that is higher than the 
consumption of the ISM components during the astration process; in fact, many of these H$_2$-dominated galaxies also contain a lot of hot X-ray emitting gas and dust grains could be destroyed by collisional sputtering 
by the hot gas atoms \citep[e.g.,][]{jones04}.

{\em ii)} The emission cross-section of dust grains associated to the molecular component could be smaller than 
that in H{\sc i}-rich environments. Indeed, a pilot study on two resolved DustPedia galaxies finds a reduced 
cross section in regions of high gas surface density \citep{clark19}. If this is confirmed, and if the grain
cross-section in molecular clouds is smaller than the THEMIS value calibrated on FIR observations of the Milky 
Way high-Galactic latitude cirrus, the current values of the dust mass for H$_2$-rich galaxies, therefore also the D/H ratios, could be 
underestimated.
We discuss this issue further in a companion paper \citep{bianchi19}. 

{\em iii)} Most of the active galaxies have high molecular fractions, as shown in Fig.~\ref{fig:fh2}. In addition,  some  H{\sc i}-deficient galaxies undergoing 
an interaction with the intercluster medium are also located in the right-hand tail of Fig.~\ref{fig:fh2}.
Active galaxies might have an H$_2$ enhancement in the center \citep[e.g.,][]{casasola08,casasola11} and the use of a uniform radial scale-length 
for the whole gaseous sample might overestimate the molecular mass and the gas molecular fraction. 
However, some of the active galaxies are also starburst and might indeed have a real enhancement of the molecular gas fraction 
while most of the dust is destroyed by the intense radiation field mentioned in $i)$. 
For H{\sc i}-deficient galaxies dust and H{\sc i} might be removed from the optical disk during the interaction with the intercluster medium 
\citep[e.g.,][]{cortese10,corbelli12}
while they often retain their molecular content or even enhance it by gas compression. 
Therefore, we expect them to have a lower dust-to-gas mass ratio if the galaxy is intrinsically H$_2$ dominated. 
As shown by \citet{stark13}, galaxies with a high molecular fraction often have a blue central region often referred to as mass-corrected blue-centeredness which is linked to enhanced star formation and external perturbations.

{\em iv)} 
The adoption of the Galactic $X_{\rm CO}$ conversion factor, both constant and metallicity-dependent, for the whole sample could also partially contribute 
to explain the decrease of DGR in some H$_2$-dominated galaxies.
As mentioned in {\em iii)}, some of these galaxies are also starburst where the molecular gas is typically warmer, denser, and with higher column densities
than in the Milky Way and in less active objects \citep[e.g.,][]{bradford03,ward03,rangwala11}.
These physical conditions of the gas produce a lower $X_{\rm CO}$ conversion factor 
\citep[$X_{\rm CO} \sim 0.4 \times 10^{20}$ cm$^{-2}$ (K~km~s$^{-1}$)$^{-1}$, e.g.,][]{downes98,Papadopoulos12,bolatto13} than the Galactic one, and
therefore a higher DGR for some galaxies located in the decreasing part of the hill-like distribution shown in Fig.~\ref{fig:hill}.

It is well known that also the presence of a companion galaxy and/or a stellar bar can affect this trend as well as the other scaling relations presented above \citep[e.g.,][and many others]{braine93,combes94,casasola04,dimatteo07,Khoperskov18,Moreno19}. 
However, a detailed analysis on these aspects is beyond the scope of this paper, and will be addressed in future publications within the DustPedia collaboration. 

To our knowledge, no simulation or theory papers make predictions for the resulting DGR versus H$_2$/H{\sc i} distribution. 
Our analysis therefore provides new constraints for cosmological models of galaxy evolution.

\section{Summary and final remarks}
\label{sec:conclusions}  
We present a study on the main ISM scaling relations for the nearby late-type galaxies of the Local Universe.
We also discuss important mass ratios.
The analysis is based on the DustPedia galaxy sample and the DustPedia dataset, including M$_{\rm d}$,  M$_{\rm star}$, oxygen abundance, and galaxy morphology. 
Molecular and atomic gas data were extracted from the literature and uniformly homogenized.
All the masses are reported to their value within $r_{25}$.
The scaling relations and mass ratios involving molecular gas mass were explored under the assumption of both a constant and a metallicity-dependent X$_{\rm CO}$.
Our analysis provides new constraints for cosmological models of galaxy evolution and a reference for high-redshift studies.
Our main results can be summarized as follows.
\begin{itemize}
\item Dust and total gas are better correlated than dust and molecular gas, or dust and atomic gas. 
Dust and atomic gas are better correlated than dust and molecular gas, and this is opposite to what is observed at smaller scales in the ISM.
All these scaling relations have a higher Pearson correlation coefficient under the assumption of a constant $X_{\rm CO}$  than under the assumption of a metallicity-dependent one.  
\item Combining H{\sc i}, H$_2$, and stellar masses, we provide general gas-star scaling relations for late-type galaxies of the Local Universe.
These gas-star scaling relations allow us to derive the mass of a gas phase (typically the H$_2$) given the mass of the other gas phase and the stellar mass for a given galaxy. 
\item We provide a characterization of the following mass ratios  as a function of morphological stage: gas fraction, molecular-to-atomic gas mass ratio, DGR, and dust-to-stellar mass ratio.
Among these mass ratios, only DGR shows a different behavior based on the assumption on $X_{\rm CO}$.  
\item Only the assumption of the metallicity-dependent $X_{\rm CO}$ is able to reproduce the expected decrease in the DGR as a function of galaxy morphology and gas-phase metallicity. 
Our analysis shows that DGR, gas-phase metallicity, and galaxy morphology are properties that are directly linked. 
Based on these findings, we recommend the use of a metallicity-dependent $X_{\rm CO}$, especially to characterize DGR.
\item We explore a novel trend between DGR  and molecular-to-atomic gas mass ratio.
The data are distributed along a hill-like curve for both assumptions on $X_{\rm CO}$, where the gas molecular fraction and the DGR show a positive correlation 
for low molecular gas fractions, while this trend breaks down for molecular-rich galaxies. 
We suggest several different scenarios to interpret this finding.
\end{itemize}

\begin{acknowledgements}
We are grateful to the anonymous referee, whose comments and suggestions improved the quality of this manuscript. 
The authors would like to thank A.~Giannetti and F.~Calura for helpful discussions.
DustPedia is a collaborative focused research project supported by the European Union under the Seventh Framework Programme (2007-2013) call (proposal no. 606824). 
The participating institutions are: Cardiff University, UK; National Observatory of Athens, Greece; Ghent University, Belgium; Universit\'{e} Paris Sud, France; 
National Institute for Astrophysics, Italy and CEA (Paris), France. 
We acknowledge funding from the INAF PRIN-SKA 2017 program 1.05.01.88.04.
We acknowledge funding from the INAF main stream 2018 program ``Gas-DustPedia: A definitive view of the ISM in the Local Universe''.
VC, SB, and EC acknowledge the support from grant PRIN MIUR 2017 - 20173ML3WW$\_$001.
This research made use of the NASA/IPAC Extragalactic Database (NED), which is operated by the Jet Propulsion Laboratory, 
California Institute of Technology, under contract with the National Aeronautics and Space Administration. 
This research has made use of the VizieR catalogue access tool, CDS, Strasbourg, France. 
The original description of the VizieR service was published in \citet{ochsenbein00}.
We acknowledge the usage of the HyperLeda database (http://leda.univ-lyon1.fr).
\end{acknowledgements}


\appendix

\section{\ion{H}{i} and CO observations}
\label{sec:hicodata}
In this section, we collect the main properties of our gas data  in three tables.
Table~\ref{tab:refs} lists references consulted for $^{12}$CO data, and Tables~\ref{tab:CO-telescopes} and ~\ref{tab:HI-telescopes} 
list telescopes and main observational properties characterizing $^{12}$CO and H{\sc i} data, respectively.

\begin{table}
\caption{\label{tab:refs} 
Sources of $^{12}$CO data collected for this work.}
\centering
\begin{tabular}{llccc}
\hline\hline
Reference                       & No. galaxies\\
\hline
\citet{aalto94} &    1 \\
\citet{Albrecht04} & 5 \\
\citet{bajaja95} &   2 \\
\citet{bettoni01} &  5\\
\citet{bettoni03} & 42 \\
\citet{boselli14} &140\\
\citet{Bourne13} &   6 \\
\citet{casasola04} &12 \\
\citet{claussen92} & 2 \\
\citet{corbelli12} & 2 \\
\citet{dahlem93} &   1 \\
\citet{elfhag96} &   2 \\
\citet{GomezdeCastro97} & 1 \\
\citet{horellou95} & 4 \\
\citet{hunt15b} &     1 \\
\citet{hunt17} &     1\\
\citet{israel92} &   1 \\
\citet{israel95} &   1 \\
\citet{kenney88} &   1 \\
\citet{kenney89} &   1 \\
\citet{kenney90} &   1 \\
\citet{kohno03} &    1 \\
\citet{kuno07} &     1 \\
\citet{leroy09} &    1\\
\citet{maiolino97} &10 \\
\citet{planesas89} & 1 \\
\citet{rahman12} &   1 \\
\citet{roberts91} &  3 \\
\citet{sage93a} &    1 \\
\citet{sage93b} &    5 \\
\citet{sakamoto06} & 1 \\
\citet{stark86} &    1 \\
\citet{Vila-Vilaro15}&1 \\
\citet{wilson12} &   4 \\
\citet{young95} &   52 \\
\citet{young11} &    1 \\
\hline
\hline
\end{tabular}
\tablefoot{
We provide the number of galaxies extracted from each reference.    
In some cases, more than one reference has been consulted to collect all needed information for a given galaxy (see main text for details).
}
\end{table}

\begin{table}
\caption{\label{tab:CO-telescopes} 
Main properties of the telescopes used for the collected $^{12}$CO data.}
\centering
\begin{tabular}{lccccc}
\hline\hline
Telescope$^{1}$ & $^{12}$CO line        & $\theta_{\rm beam}$$^{2}$ & Units & Jy/K  \\
        
                                &                               & [$^{\prime\prime}$] & [K, Jy, M$_\odot$]\\
\hline
\textbf{Single-dish} \\
BELL            & (1--0) & 100  & $T^*_{\rm mb}$        & 108 \\
FCRAO           & (1--0) & 45   & $T^*_{\rm A}$ & 42            \\ 
IRAM-30m        & (1--0) & 22   & $T^*_{\rm mb}$        & 4.8\\
IRAM-30m        & (2--1) & 11   & $T^*_{\rm mb}$        & 5.3 \\
JCMT            & (2--1) & 22   & $T^*_{\rm A}$         & 27 \\
JCMT            & (3--2) & 14.5         & $T^*_{\rm A}$         & 33 \\
NRAO-12m        & (1--0) & 55   & $T^*_{\rm R}$         & 35\\ 
OSO             & (1--0) & 34   & $T^*_{\rm mb}$        & 12 \\
SEST            & (1--0) & 44   & $T^*_{\rm mb}$        & 27\\
SEST            & (2--1) & 24   & $T^*_{\rm mb}$        & 41 \\
\hline
\textbf{Interferometer} \\
NMA             & (1--0) & --   & $S$$^{3}$             & --\\
CARMA           & (1--0) & --   & M(H$_2$)$^{4}$        & --\\
SMA                     & (1--0) & --   & M(H$_2$)$^{4}$        & --\\ 
\hline
\hline
\end{tabular}
\tablefoot{
$^{1}$~BELL is the Bell 7m telescope \citep[][]{chu78}, 
FCRAO the 14m Five College Radio Astronomical Observatory,  
IRAM-30m the 30m antenna of the \textit{Institut de Radioastronomie Millim{\'e}trique}, 
NRAO-12m the 12m National Radio Astronomy Observatory,
OSO the 25m Onsala Space Observatory, the Swedish National Facility for Radio Astronomy,
SEST the 15m Swedish-ESO Submillimetre Telescope,
JCMT the 15m James Clerk Maxwell Telescope,
NMA is the Nobeyama Millimeter Array, an interferometer of six 10m antennas,
CARMA, Combined Array for Research in Millimeter-wave Astronomy, an interferometer of six 10.4m, nine 6.1m, and eight 3.5m antennas,
SMA, Submillimeter Array, an interferometer of eight 6m antennas.      
$^{2}$~$\theta_{\rm beam}$ is the FHWM of the listed single-dish antennas (primary beam), no information is listed for the interferometers since 
those observations cover fields of view larger than the primary beam of the single antennas composing the interferometer.     
$^{3}$~Flux provided in Jy.
$^{4}$~H$_2$ mass value transformed into total flux in Jy~km~s$^{-1}$, by adopting the distance and the $X_{\rm CO}$ conversion factor used in the original reference. 
}
\end{table}

\begin{table}
\caption{\label{tab:HI-telescopes} 
Telescopes and beams used for the collected H{\sc i} data.}
\centering
\begin{tabular}{lccccc}
\hline\hline
Telescope       & $\theta_{\rm beam}$ \\
                                & [$^{\prime}$] \\
\hline
\textbf{Single-dish} \\
Arecibo         & 3  \\
Effelsberg              & 9 \\
Green Bank 42m & 21 \\
Green Bank 91m & 10 \\
Jodrell Bank    & 11\\
Nancay          & 9 \\
Parkes-64m      & 14\\
Westerbork      & 43 \\
76m-Lovell Telescope    & 12 \\
\hline
\hline
\end{tabular}
\tablefoot{
We do not provide $\theta_{\rm beam}$ for interferometric observations (VLA), since they cover the entire optical radius of the observed galaxies 
($r_{25}$).
}
\end{table}

\section{Caveats}
\label{sec:caveats}

In this section, we  list  possible  caveats  and  systematic  uncertainties associated with the analysis and results presented above. 
These can arise from our simplified assumptions that are not able to completely represent the physics occurring in the ISM
and from the adopted methodologies to derive the various physical quantities.

\subsection{Caveats concerning the derivation of the $^{12}$CO emission within the optical disk and of the molecular gas mass}  
\label{sec:caveats-xco}
Most of the collected $^{12}$CO data are observations of one position towards the galaxy center (see Sect.~\ref{sec:co}), 
from which we derived the corresponding $^{12}$CO flux within $r_{25}$ assuming an 
exponential radial and vertical distribution, central peak, with a scale-length of $h_{\rm CO}/r_{\rm 25} = 0.17 \pm 0.03$
\citep[][]{casasola17} and a scale-height of $z_{\rm CO}/r_{\rm 25} = 1/100$ \citep[][see Sect.~\ref{sec:derco}]{boselli14}.
The adopted value for the molecular gas scale-length, although consistent with those derived in other works, may not be fully representative 
of a large galaxy sample, spanning a wide range of galaxy properties, such as that presented in this work.      
As already discussed in Sect.~\ref{sec:derco}, the exponential decline of $^{12}$CO is the most common trend in galaxies but all distributions have been observed. 
The value assumed for the scale-height of $^{12}$CO is expected to be subject to a larger uncertainty since its determination is based 
on a small number of galaxies (see Sect.~\ref{sec:derco}).
Despite the uncertainty on $h_{\rm CO}/r_{\rm 25}$, we preferred to take into account the vertical distribution of the molecular gas.
We also checked the location of sample galaxies to which no radial corrections have been applied (e.g., NGC~1068, NGC~4013) in the main plots 
of our work, such as Figs.~\ref{fig:logd-g-Z} and \ref{fig:hill}.
These galaxies cover different regions of the plots confirming that the resulting trends are real and not driven by assumptions on the $^{12}$CO distribution
in the galactic disk.      
We also stress that \citet{corbelli12} studied a sample of 18 galaxies mapped in $^{12}$CO together with a sample 
of 17 galaxies where the total $^{12}$CO flux was determined in a similar way to in our study.
\citet{corbelli12} did not report differences between the two samples in terms of relationship between dust and molecular or total gas content 
noticing that galaxies with total derived $^{12}$CO flux are very well represented by exponential functions \citep[see also][]{kuno07}. 
  
We derived M(H$_2$) by adopting both a constant and a metallicity-dependent $X_{\rm CO}$ conversion factor.
Both assumptions suffer from systematic uncertainties which propagate to the M$_{{\rm{H}}_2}$ estimate.    
The choice of a constant $X_{\rm CO}$ is supported by studies of  ``normal'' galaxies which return similar values in Milky Way-like disks, 
but with greater scatter and systematic uncertainty.   
However, departures from the Galactic conversion factor are both observed and expected, and
estimates of ``typical'' values of $X_{\rm CO}$ in the Milky Way and other spiral galaxies range
from $\sim$1.5 $\times$ 10$^{20}$ cm$^{-2}$ (K km s$^{-1}$)$^{-1}$ to
$\sim$4 $\times$ 10$^{20}$ cm$^{-2}$ (K km s$^{-1}$)$^{-1}$ (see references quoted in Sect.~\ref{sec:masses}).
We decided to adopt a constant $X_{\rm CO}$ for the sake of uniformity and simplicity, as 
suggested in the review of \citet{bolatto13} in the absence of a detailed characterization 
of $X_{\rm CO}$ in a given galaxy and as typically done in the major portion of studies aimed at determining molecular 
gas mass from $^{12}$CO line intensities.
However, within the DustPedia collaboration we also uniformly derived the oxygen abundances of the entire DustPedia galaxy sample 
which allowed us to take into account the dependence of $X_{\rm CO}$ on the metallicity.
This clearly adds further uncertainties due to the method used to derive the metallicity (see later Sect~\ref{sec:caveats-metalicity}) 
and the chosen $X_{\rm CO}$-metallicity calibration (see Sect~\ref{sec:metallicity}).    
    
Galaxies with available $^{12}$CO data coming from $^{12}$CO(2--1) or $^{12}$CO(3--2) emission lines 
could be affected by an uncertainty due to the assumption of a mean ratio between $^{12}$CO emission lines, and not
determined for a given galaxy. 
Without precise knowledge of $^{12}$CO line ratios for all galaxies, we indeed adopted typical ratios present in the literature 
(see Sect.~\ref{sec:gasdata}).   
However, we signal that only 4.7$\%$ of the  $^{12}$CO data come from $^{12}$CO(2--1) and $^{12}$CO(3--2) transitions. 

Data of $^{12}$CO coming from interferometric observations 
(see Table~\ref{tab:CO-telescopes}) 
can also suffer from an additional uncertainty. 
An interferometer is limited by the minimum spacing of its antennas. 
Because two antennas cannot be placed closer than some minimum distance ($D_{\rm min}$), signals on spatial scales larger than
some size ($\propto \lambda/D_{\rm min}$) will be attenuated. 
This effect, called the ``missing flux'' problem, is typically resolved by using single-dish observations to compute short spacings 
and complete the interferometric measurements \citep[see e.g.,][as examples of $^{12}$CO short-spacing corrections]{casasola08,casasola10,casasola11,combes09,vanderlaan11}.
Unfortunately the interferometric $^{12}$CO observations that we collected in this work are not corrected for short spacings, but they only represent 
$\sim$1.2$\%$ of $^{12}$CO data.

\subsection{Caveats concerning the derivation of the H{\sc i} emission within the optical disk}  
\label{sec:caveats-hi}
As presented in Sect.~\ref{sec:hi}, we derived the H{\sc i} emission within the optical disk
following the method of \citet{wang14}, who developed an empirical model to describe the 
observed exponential H{\sc i} profiles as a function of radius in the 
outer parts of the galaxy with 
a depression towards the center.
\citet{wang14} have also compared their results with existing smoothed particle hydrodynamical
simulations and semi-analytic models of disk galaxy formation in a $\Lambda$ cold dark matter Universe.
Both the hydro simulations and the semi-analytic models are able to reproduce the H{\sc i} surface
density profiles and the H{\sc i} size-mass relation without further tuning of the simulation and
model inputs.
The ``Universal'' outer disk profiles in the semi-analytic methods  
originate from the combination of the assumption that infalling gas has an exponential profile and of the inside-out
growth of disks \citep[e.g.,][]{kauffmann96,dutton07,fu09}.
The limit of these findings is that the assumption that gas accreted from the halo is
distributed over the disk with an exponential profile has no a priori
physical justification.
As pointed out by \citet{wang14},  the fact that the hydrodynamical simulations, 
at least those presented by \citet{wang14}, have outer disks that agree so well with
the observed ``Universal'' outer exponential profiles seen in the data is therefore somewhat incomprehensible.     
The explanation likely lies in the complex interplay between the infall
of new gas, star formation, supernova feedback, and gas inflow
processes occurring in the simulation.

\subsection{Caveats concerning the gas-phase metallicity derivation}
\label{sec:caveats-metalicity}
As mentioned in Sect.~\ref{sec:metallicity}, we adopted gas-phase metallicity determinations from 
the empirical calibration N2 of \citet{pettini04}.
These metallicities were extracted from \citet{devis19}, who compared 
the results from different empirical and theoretical methods to understand any systematic differences 
that may result from using various methods.
There are limitations linked to the type of metallicity calibration. For example, 
the empirical calibrations, like the N2 one, are only valid for the same range of excitation and
metallicity as the H{\sc ii} regions that were used to build the calibration.   
Since they are determined assuming an electron temperature, 
these methods may systematically underestimate the true metallicity if there are temperature inhomogeneities in a galaxy
\citep[e.g.,][]{devis19}.

\end{document}